\documentclass[journal]{IEEEtran}

\usepackage{epsfig}
\usepackage{latexsym}
\usepackage{url}
\usepackage{float}
\usepackage[hypertexnames=false,hyperfootnotes=false]{hyperref}
\usepackage{texnansi}
\usepackage{color}
\usepackage{tikz}
\usepackage{subfigure}
\usepackage{afterpage}
\usepackage{enumerate}
\usepackage{algorithm}
\usepackage{algorithmic}
\usepackage[normalem]{ulem}
\usepackage{lmodern}
\usepackage{ifthen}
\usepackage[cmex10]{amsmath}
\usepackage{amsthm}
\usepackage{amssymb}
\usepackage{amsfonts}
\usepackage{cite}
\usepackage{pgfplots}

\newcommand{\field}[1]{\ensuremath{\mathbb{#1}}}
\newcommand{\R}{\ensuremath{\field{R}}} 
\newcommand{\I}[1]{\ensuremath{\mathbb{I}_{\left\{#1\right\}}}} 
\newcommand{\Inb}[1]{\ensuremath{\mathbb{I}_{#1}}} 
\newcommand{\tends}{\ensuremath{\rightarrow}} 
\newcommand{\E}{\ensuremath{\mathsf{E}}} 
\newcommand{\defeq}{\ensuremath{\triangleq}}
\newcommand{\Ascr}{\ensuremath{\mathcal A}}
\newcommand{\Bscr}{\ensuremath{\mathcal B}}

\newcommand{\Fscr}{\ensuremath{\mathcal F}}
\newcommand{\Gscr}{\ensuremath{\mathcal G}}

\newcommand{\Iscr}{\ensuremath{\mathcal I}}

\newcommand{\Mscr}{\ensuremath{\mathcal M}}

\newcommand{\Tscr}{\ensuremath{\mathcal T}}

\newcommand{\Zscr}{\ensuremath{\mathcal Z}}

\DeclareMathOperator*{\argmin}{\mathrm{argmin}}

\newtheorem{assumption}{Assumption}
\newtheorem{definition}{Definition}
\newtheorem{example}{Example}
\newtheorem{theorem}{Theorem}

\newtheorem{lemma}{Lemma}

\usetikzlibrary{arrows,patterns,plotmarks}
\tikzstyle{every picture} += [>=stealth]
\newcommand{\emailhref}[1]{\href{mailto:#1}{\tt #1}} 
\provideboolean{fastcompile}
\newcommand{\hidefastcompile}[1]{\ifthenelse{\boolean{fastcompile}}{}{#1}}

\newcommand{\A}{\field{A}}
\newcommand{\X}{\field{X}}
\newcommand{\Y}{\field{Y}}
\newcommand{\Sf}{\field{S}}
\newcommand{\TV}{\mbox{\rm TV}} 

\newtheorem*{le:tvrate}{Lemma~\ref{le:tvrate}}
\newtheorem*{ex:rps}{Example~\ref{ex:rps}}

\newcommand{\GETS}{\leftarrow}

\begin{document}

\title{Universal Reinforcement Learning}

\author{Vivek~F.~Farias,~\IEEEmembership{Member,~IEEE,}
  Ciamac~C.~Moallemi,~\IEEEmembership{Member,~IEEE,}
  Benjamin~Van~Roy,~\IEEEmembership{Senior Member,~IEEE,}
  Tsachy~Weissman,~\IEEEmembership{Senior Member,~IEEE}%
  \thanks{Manuscript received July 20, 2007; revised June 8, 2009. The first
    author was supported by a supplement to NSF Grant ECS-9985229 provided by
    the MKIDS Program. The second author was supported by a Benchmark Stanford
    Graduate Fellowship.}%
  \thanks{V. F. Farias is with the Sloan School of Management, Massachusetts
    Institute of Technology, Cambridge, MA, 02139 USA (e-mail:
    \emailhref{vivekf@mit.edu}.)}%
  \thanks{C. C. Moallemi is with the Graduate School of Business, Columbia
    University, New York, NY, 10027 USA (e-mail:
    \emailhref{ciamac@gsb.columbia.edu}).}%
  \thanks{B. Van Roy is with the Departments of Management Science \&
    Engineering and Electrical Engineering, Stanford University, Stanford, CA
    94305 USA (e-mail: \emailhref{bvr@stanford.edu}).}%
  \thanks{T. Weissman is with the Department of Electrical Engineering,
    Stanford University, Stanford, CA 94305 USA (e-mail:
    \emailhref{tsachy@stanford.edu}).}%
}

\markboth{IEEE Transactions on Information Theory}%
{Farias \MakeLowercase{\textit{et al.}}: Universal Reinforcement Learning}

\maketitle

\begin{abstract}
  We consider an agent interacting with an unmodeled environment.  At each
  time, the agent makes an observation, takes an action, and incurs a cost.
  Its actions can influence future observations and costs.  The goal is to
  minimize the long-term average cost.  We propose a novel algorithm, known as
  the active LZ algorithm, for optimal control based on ideas from the
  Lempel-Ziv scheme for universal data compression and prediction.  We
  establish that, under the active LZ algorithm, if there exists an integer
  $K$ such that the future is conditionally independent of the past given a
  window of $K$ consecutive actions and observations, then the average cost
  converges to the optimum.  Experimental results involving the game of
  Rock-Paper-Scissors illustrate merits of the algorithm.
\end{abstract}

\begin{IEEEkeywords}
  Lempel-Ziv, context tree, optimal control, reinforcement learning, dynamic
  programming, value iteration.
\end{IEEEkeywords}

\section{Introduction}

\IEEEPARstart{C}{onsider} an agent that, at each integer time $t$, makes an
observation $X_t$ from a finite observation space $\X$, and takes an action
$A_t$ selected from a finite action space $\A$. The agent incurs a bounded
cost $g(X_t,A_t,X_{t+1}) \in [-g_{\max},g_{\max}]$.  The goal is to minimize
the long-term average cost
\[
\limsup_{T \rightarrow \infty}\  
\E\left[
  \frac{1}{T}
  \sum_{t=1}^{T} g(X_t,A_t,X_{t+1})
\right].
\]
Here, the expectation is over the randomness in the $X^t$
process\footnote{ For a sequence such as $\{X_t\}$, $X_s^t$ denotes
  the vector $(X_s, \ldots, X_t)$. We also use the notation
  $X^t=X_1^t$.}, and, at each time $t$, the action $A_t$ is selected
as a function of the prior observations $X^t$ and the prior actions
$A^{t-1}$.

We will propose a general action-selection strategy called the \emph{active LZ
  algorithm}.  In addition to the new strategy, a primary contribution of this
paper is a theoretical guarantee that this strategy attains optimal average
cost under weak assumptions about the environment.  The main assumption is
that there exists an integer $K$ such that the future is conditionally
independent of the past given a window of $K$ consecutive actions and
observations.  In other words,
\begin{equation}\label{eq:transprob}
\Pr\left(\left.X_{t}=x_t\right|\Fscr_{t-1}\right)
=P\big(x_t\big|X^{t-1}_{t-K},A^{t-1}_{t-K}\big),
\end{equation}
where $P$ is a transition kernel and $\Fscr_t$ is the $\sigma$-algebra
generated by $(X^t,A^t)$. We are particularly interested in situations
where neither $P$ nor even $K$ are known to the agent. That is, where
there is a finite but unknown dependence on history.

Consider the following examples, which fall into the above formalism.
\begin{example}[Rock-Paper-Scissors]\label{ex:rps}
  Rock-Paper-Scissors is a two-person, zero-sum matrix game that has a rich
  history as a reinforcement learning problem. The two players play a series
  of games indexed by the integer $t$. Each player must generate an
  action---rock, paper, or scissors---for each game.  He then observes his
  opponent's hand and incurs a cost of $-1$, $1$, or $0$, depending on
  whether the pair of hands results in a win, loss, or draw.  The game is
  played repeatedly and the player's objective is to minimize the average
  cost.
  
  Define $X_{t}$ to be the opponent's choice of action in game $t$,
  and $A_{t-1}$ to be the player's choice of action in game $t$. The
  action and observation spaces for this game are 
  \[
  \A \defeq \X
  \defeq \{ \text{rock}, \text{paper}, \text{scissors} \}.
  \] 
  Identifying these with the integers $\{1,2,3\}$, the cost function is
  \[
  g(x_t,a_t,x_{t+1}) \defeq
  \begin{bmatrix}
    0 & 1 & -1 \\
    -1 & 0 & 1 \\
    1 & -1 & 0 
  \end{bmatrix}_{x_{t+1},a_t}.
  \]
  Assuming that the opponent uses a mixed strategy that depends only
  on information from the last $K-1$ games, such a strategy defines
  a transition kernel $P$ over the opponent's play $X_{t}$ in game
  $t$ of the form \eqref{eq:transprob}. (Note that such a $P$ has
  special structure in that, for example, it has no dependence on
  the player's action $A_{t-1}$ in game $t$, since this is unknown
  to the opponent until after game $t$ is played.) Then, the problem
  of finding the optimal strategy against an unknown, finite-memory
  opponent falls within our framework.
\end{example}

\begin{example}[Joint Source-Channel Coding with a Fixed Decoder] Let $\Sf$
  and $\Y$ be finite source and channel alphabets, respectively. Consider a
  sequence of symbols $\{S_t\}$ from the source alphabet $\Sf$ which are to
  be encoded for transmission across a channel. Let $Y_t\in\Y$ represent the
  choice of encoding at time $t$, and let $\hat{Y}_t\in\Y$ be the symbol
  received after corruption by the channel. We will assume that this channel
  has a finite memory of order $M$. In other words, the distribution of
  $\hat{Y}_t$ is a function of $Y^t_{t-M+1}$. For all times $t$, let $d:\
  \Y^{L} \rightarrow \Sf$ be some fixed decoder that decodes the symbol at
  time $t$ based on the most recent $L$ symbols received
  $\hat{Y}^t_{t-L+1}$. Given a single letter distortion measure $\rho:\ \Sf
  \times \Sf \rightarrow \R$, define the expected distortion at time $t$ by
  \begin{multline*}
    \lefteqn{g(s_t,y^t_{t-L-M+2})} \\
     \defeq
     \E\left[
       \left.
         \rho\left(d(\hat{Y}^t_{t-L+1}), s_t \right)\ 
       \right|\ Y^t_{t-L-M+2}=y^t_{t-L+M+2}
     \right].
   \end{multline*}
   The optimization problem is to find a sequence of functions $\{ \mu_t
  \}$, where each function $\mu_t: \X^t \rightarrow \A$ specifies an
  encoder at time $t$, so as to minimize the long-term average distortion
  \[
  \limsup_{T\tends\infty}\ 
  \E_\mu\left[
    \frac{1}{T} 
    \sum_{t=1}^T
    g(S_t,Y^t_{t-L-M+2})
  \right].
  \]
  Assume that the source is Markov of order $N$, but that both the transition
  probabilities for the source and the order $N$ are unknown. Setting
  $K=\max(L+M-1,N)$, define the observation at time $t$ to be the vector
  $X_t=(S_t,Y^{t-1}_{t-L-M+2})$ and the action at time $t$ to be
  $A_t=Y_t$. Then, optimal coding problem at hand falls within our framework
  (cf. \cite{Teneketzis06} and references therein).
\end{example}

With knowledge of the kernel $P$ (or even just the order of the kernel, $K$),
solving for the average cost optimal policy in either of the examples above
via dynamic programming methods is relatively straightforward. This paper
develops an algorithm that, \emph{without knowledge of the kernel or its
  order}, achieves average cost optimality. The active LZ algorithm we develop
consists of two broad components. The first is an efficient data structure, a
context tree on the joint process $(X^t,A^{t-1})$, to store information
relevant to predicting the observation at time $t+1$, $X_{t+1}$, given the
history available up to time $t$ and the action selected at time $t$,
$A_t$. Our prediction methodology borrows heavily from the Lempel-Ziv
algorithm for data compression \cite{ziv78compression}. The second component
of our algorithm is a dynamic programming scheme that, given the probabilistic
model determined by the context tree, selects actions so as to minimize costs
over a suitably long horizon. Absent knowledge of the order of the kernel,
$K$, the two tasks above---building a context tree in order to estimate the
kernel, and selecting actions that minimize long-term costs---must be done
continually in tandem which creates an important tension between `exploration'
and `exploitation'. In particular, on the one hand, the algorithm must select
actions in a manner that builds an accurate context tree. On the other hand,
the desire to minimize costs naturally restricts this selection. By carefully
balancing these two tensions our algorithm achieves an average cost equal to
that of an optimal policy with full knowledge of the kernel $P$.

Related problems have been considered in the literature. Kearns and Singh
\cite{kearns98nearoptimal} present an algorithm for reinforcement learning in
a Markov decision process.  This algorithm can be applied in our context when
$K$ is known, and asymptotic optimality is guaranteed.  More recently,
Even-Dar et al.\ \cite{Kakade05} present an algorithm for optimal control of
partially observable Markov decision processes, a more general setting than
what we consider here, and are able to establish theoretical bounds on
convergence time. The algorithm there, however, seems difficult and
unrealistic to implement in contrast with what we present here. Further, it
relies on knowledge of a constant related to the amount of time a `homing'
policy requires to achieve equilibrium. This constant may be challenging to
estimate.

Work by de Farias and Megiddo \cite{deFarias04} considers an optimal
control framework where the dynamics of the environment are not known
and one wishes to select the best of a finite set of `experts'. In
contrast, our problem can be thought of as competing with the set of
all possible strategies.  The prediction problem for loss functions
with memory and a Markov-modulated source considered by 
Merhav et al.\ \cite{Merhav02} is essentially a Markov decision problem as the
authors point out; again, in this case, knowing the structure of the
loss function implicitly gives the order of the underlying Markov
process.

The active LZ algorithm is inspired by the Lempel-Ziv algorithm.  This
algorithm has been extended to address many problems, such as prediction
\cite{Merhav98,Jacquet02} and filtering \cite{Merhav02}. In almost all cases,
however, future observations are not influenced by actions taken by the
algorithm. This is in contrast to the active LZ algorithm, which proactively
anticipates the effect of actions on future observations. An exception is the
work of Vitter and Krishnan \cite{vitter96optimal}, which considers cache
pre-fetching and can be viewed as a special case of our formulation.

The Lempel-Ziv algorithm and its extensions revolve around a context tree data
structure that is constructed as observations are made.  This data structure is
simple and elegant from an implementational point of view. The use of this data
structure in reinforcement learning represents a departure from representations
of state and belief state commonly used in the reinforcement learning
literature.  Such data structures have proved useful in representing experience
in algorithms for engineering applications ranging from compression to
prediction to denoising.  Understanding whether and how some of this value can
be extended to reinforcement learning is the motivation for this paper.

The remainder of this paper is organized as follows. In
Section~\ref{sec:clsol}, we formulate our problem precisely. In
Section~\ref{sec:unischeme}, we present our algorithm and provide
computational results in the context of the rock-paper-scissors
example. Our main result, as stated in Theorem~\ref{th:opt} in
Section~\ref{sec:analysis}, is that the algorithm is asymptotically
optimal. Section~\ref{sec:future} concludes.

\section{Problem Formulation}\label{sec:clsol}

Recall that we are endowed with finite action and observation spaces $\A$ and
$\X$, respectively, and we have 
\[
\Pr\left(\left.X_{t}=x_t\right|\Fscr_{t-1}\right)
=P\big(x_t\big|X^{t-1}_{t-K},A^{t-1}_{t-K}\big),
\]
where $P$ is a stochastic transition kernel.  A \emph{policy}
$\mu$ is a sequence of mappings $\{ \mu_t \}$, where for each time $t$ the map
$\mu_t:\ \X^t\times\A^{t-1} \rightarrow \A$ determines which action shall be
chosen at time $t$ given the history of observations and actions observed up
to time $t$. In other words, under policy $\mu$, actions will evolve according
to the rule
\[
A_t = \mu_t(X^t,A^{t-1}).
\]
We will call a policy $\mu$ stationary if 
\[
\mu_t(X^t,A^{t-1}) =
\mu(X^t_{t-K+1},A^{t-1}_{t-K+1}),\quad\text{for all $t\geq K$},
\]
for some function $\mu:\ \X^K\times\A^{K-1}\rightarrow \A$.  Such a policy
selects actions in a manner that depends only one the current observation $X_t$
and the observations and actions over the most recent $K$ time steps. It is
clear that for a fixed stationary policy $\mu$, the observations and actions for
time $t\geq K$ evolve according to a Markov chain on the finite state space
$\X^K \times \A^{K-1}$.  Given an initial state $(x^{K},a^{K-1})$, we can define
the average cost associated with the stationary policy $\mu$ by
\begin{multline*}
 \lambda_\mu(x^K,a^{K-1}) 
\\
\defeq
\lim_{T\rightarrow\infty}
\E_\mu\left[\left.
\frac{1}{T}
\sum_{t=1}^T g(X_t,A_t,X_{t+1})\ 
\right|\ 
x^K,a^{K-1}
\right].
\end{multline*}
Here, the expectation is conditioned on the initial state $(X^K,A^{K-1}) =
(x^K,a^{K-1})$.  Since the underlying state-space, $\X^K \times \A^{K-1}$, is
finite, the above limit always exists \cite[Proposition
4.1.2]{BertsekasDP}. Since there are finitely many stationary policies, we can
define the optimal average cost over stationary policies by
\[
\lambda^*(x^K,a^{K-1}) \defeq
\min_{\mu}\ 
\lambda_\mu(x^K,a^{K-1}),
\]
where the minimum is taken over the set of all stationary policies.  Again,
because of the finiteness of the underlying state space, $\lambda^*$ is also the
optimal average cost that can be achieved using any policy, stationary or
not. In other words,
\begin{equation}\label{eq:avgcost}
\begin{split}
\lefteqn{\lambda^*(x^K,a^{K-1})}
\\
& = 
\inf_{\nu}\ 
\limsup_{T\rightarrow\infty}\ 
\E_\nu\left[\left.
\frac{1}{T}
\sum_{t=1}^T g(X_t,A_t,X_{t+1})\ 
\right|\ 
x^K,a^{K-1}
\right],
\end{split}
\end{equation}
where the infimum is taken over the set of \emph{all} policies \cite[Proposition 4.1.7]{BertsekasDP}.

We next make an assumption that will enable us to streamline our analysis in
subsequent sections.
\begin{assumption}
\label{as:eqavgcost}
The optimal average cost is independent of the initial state. That is,
there exists a constant $\lambda^*$ so that
\[
\lambda^*(x^K,a^{K-1}) = \lambda^*,
\quad\forall\ (x^K,a^{K-1})\in\X^{K}\times\A^{K-1}.
\]
\end{assumption}
The above assumption is benign and is satisfied for any strictly positive
kernel $P$, for example. More generally, such an assumption holds for the
class of problems satisfying a `weak accessibility' condition (see Bertsekas
\cite{BertsekasDP} for a discussion of the structural properties of average
cost Markov decision problems). In the context of our problem, it is difficult
to design controllers that achieve optimal average cost in the absence of such
an assumption. In particular, if there exist policies under which the chain
has multiple recurrent classes, then the optimal average cost may well depend
on the initial state and actions taken very early on might play a critical
role in achieving this performance. We note that in such cases the assumption
above (and our subsequent analysis) is valid for the recurrent class our
controller eventually enters.

If the transition kernel $P$ (and, thereby, $K$) were known, dynamic programming
is a means to finding a stationary policy that achieves average cost
$\lambda^*$. One approach would be to find a solution $J:\
\X^K\times\A^{K-1}\rightarrow\R$ to the discounted Bellman equation
\begin{equation}\label{eq:bellman}
\begin{split}
\lefteqn{J(x^K,a^{K-1}) } \\
&
=
\min_{a_K}\ 
\sum_{x_{K+1}}\ 
P(x_{K+1}|x^K,a^K)
\\
&
\qquad\qquad\qquad\times
\big[
g(x_k,a_k,x_{K+1}) 
+ 
\alpha
J( x_2^{K+1}, a_2^{K})
\big],
\end{split}
\end{equation}
for all $(x^K,a^{K-1})\in\X^{K}\times\A^{K-1}$.  Here, $\alpha\in (0,1)$ is a
discount factor. If the discount factor alpha is chosen to be sufficiently
close to 1, a solution $J_\alpha^*$ (known as the \emph{cost-to-go function})
to the Bellman equation can be used to define an optimal stationary policy for
the original, average-cost problem \eqref{eq:avgcost}. In particular, for all
$(x^K,a^{K-1})\in\X^{K}\times\A^{K-1}$, define the set
$\Ascr_\alpha^*(x^K,a^{K-1})$ of $\alpha$-discounted optimal actions to be the
set of minimizers to the optimization program
\begin{equation}\label{eq:optact}
\begin{split}
\lefteqn{\min_{a_K}\ 
\sum_{x_{K+1}}
P(x_{K+1}|x^K,a^K)}
\\
&
\qquad\qquad\quad\times
\big[
g(x_K,a_K,x_{K+1}) 
+ 
\alpha
J_\alpha^*( x_2^{K+1}, a_2^{K})
\big].
\end{split}
\end{equation}
At a give time $t$, $\Ascr_\alpha^*(X^t_{t-K+1},A^{t-1}_{t-K+1})$ is the set
of actions obtained acting greedily with respect to $J^*_\alpha$. These
actions seek to minimize the expected value of the immediate cost
$g(X_t,A_t,X_{t+1})$ at the current time, plus a continuation cost, which
quantifies the impact of the current decision on all future costs, and is
captured by $J^*_\alpha$.

If $\alpha$ is sufficiently close to 1, and $\mu^*$ is a policy such that for
$t\geq K$,
\begin{equation}\label{eq:optact2}
\mu^*_t(X^t,A^{t-1}) 
\in
\Ascr_\alpha^*(X^t_{t-K+1},A^{t-1}_{t-K+1}),
\end{equation}
then, $\mu^*$ will achieve the optimal average cost $\lambda^*$. Such a policy
$\mu^*$ is sometimes called a \emph{Blackwell optimal policy}
\cite{BertsekasDP}.

We return to our example of the game of Rock-Paper-Scissors, to illustrate the
above approach.

\begin{ex:rps}[Rock-Paper-Scissors] Given knowledge of the opponent's
  (finite-memory) strategy and, thus the transition kernel $P$, the Bellman
  equation \eqref{eq:bellman} can be solved for the optimal cost-to-go
  function $J^*_\alpha$. Then, an optimal policy for the player would be,
  for each game $t+1$, to select an action $A_t$ according to
  \eqref{eq:optact}--\eqref{eq:optact2}. This action is a function of the
  entire history of game play only through the sequence
  $(X^t_{t-K+1},A^{t-1}_{t-K+1})$ of recent game play. The action is
  selected by optimally accounting for both the expected immediate cost
  $g(X_t,A_t,X_{t+1})$ of the game at hand, and the impact of the choice of
  action towards all future games (through the cost-to-go function
  $J_\alpha^*$).
\end{ex:rps}

\section{A Universal Scheme}\label{sec:unischeme}

Direct solution of the Bellman equation \eqref{eq:bellman} requires knowledge
of the transition kernel $P$. Algorithm~\ref{alg:lzlearn}, the \emph{active LZ
  algorithm}, is a method that requires no knowledge of $P$, or even of
$K$. Instead, it simultaneously estimates a probabilistic model for the
evolution of the system and develops an optimal control for that model, along
the course of a single system trajectory. At a high-level, the two critical
components of the active LZ algorithm are the estimates $\hat{P}$ and
$\hat{J}$. $\hat{P}$ is our estimate of the true kernel $P$. This estimate is
computed using variable length contexts to dynamically build higher order
models of the underlying process, in a manner reminiscent of the Lempel-Ziv
scheme used for universal prediction. $\hat{J}$ is the estimate to the optimal
cost-to-go function $J^*_\alpha$ that is the solution to the Bellman equation
\eqref{eq:bellman}. It is computed in a fashion similar to the value iteration
approach to solving the Bellman equation equation (see
\cite{BertsekasDP}). Given the estimates $\hat{P}$ and $\hat{J}$, the
algorithm randomizes to strike a balance selecting actions so as to improve
the quality of the estimates (exploration) and acting greedily with respect to
the estimates so as to minimize the costs incurred (exploitation).

The active LZ algorithm takes as inputs a discount factor
$\alpha\in (0,1)$, sufficiently close to $1$, and a sequence of
exploration probabilities $\{ \gamma_t \}$. The algorithm proceeds as
follows: time is parsed into intervals, or `phrases', with the
property that if the $c$th phrase covers the time intervals $\tau_c
\leq t \leq \tau_{c+1}-1$, then the observation/action sequence
$(X_{\tau_c}^{\tau_{c+1}-1},A_{\tau_c}^{\tau_{c+1}-2})$ will not have
occurred as the prefix of any other phrase before time $\tau_c$. 

\begin{algorithm*}
\caption{The active LZ algorithm, a Lempel-Ziv inspired algorithm for learning.}
\label{alg:lzlearn}
\begin{algorithmic}[1]
\REQUIRE a discount factor $\alpha\in (0,1)$ and a sequence of exploration probabilities $\{\gamma_t\}$
\STATE $c \GETS 1$
\COMMENT{the index of the current phrase}
\STATE $\tau_c \GETS 1$
\COMMENT{start time of the $c$th phrase}
\STATE $N(\cdot)\GETS 0$
\COMMENT{initialize context counts}
\STATE $\hat{P}(\cdot) \GETS 1/|\X|$\label{algl:pinit}
\COMMENT{initialize estimated transition probabilities}
\STATE $\hat{J}(\cdot)\GETS 0$\label{algl:jinit}
\COMMENT{initialize estimated cost-to-go values}
\FOR {each time $t$}
\STATE observe $X_t$
\IF[are in a context that we have seen before?]
{$N(X_{\tau_c}^t,A_{\tau_c}^{t-1}) > 0$}
\STATE with probability $\gamma_t$, pick $A_t$ uniformly over $\A$
\label{algl:explore}
\COMMENT{explore independent of history}
\STATE with remaining probability, $1-\gamma_t$, pick $A_t$ greedily according to $\hat{P}$, $\hat{J}$:
\label{algl:exploit}
\[
A_t \in 
\argmin_{a_t}
\sum_{x_{t+1}}
\hat{P}\big(x_{t+1}|X_{\tau_c}^t,(A_{\tau_c}^{t-1},a_t)\big)
\left[
g(X_t,a_t,x_{t+1})+
\alpha
\hat{J}\big((X_{\tau_c}^t,x_{t+1}),(A_{\tau_c}^{t-1},a_t)\big)
\right]
\]
\COMMENT{exploit by picking an action greedily}
\ELSE[we are in a context not seen before]\label{algl:backstart}
\STATE pick $A_t$ uniformly over $\A$
\FOR[traverse backward through the current context]{ $s$ with $\tau_c \leq s \leq t$, in decreasing order }
\STATE update context count: $N(X_{\tau_c}^s,A_{\tau_c}^{s-1}) \GETS N(X_{\tau_c}^s,A_{\tau_c}^{s-1}) + 1$
\STATE update probability estimates: for all $x_s\in\X$\label{algl:pupdate}
\[
\hat{P}(x_s|X_{\tau_c}^{s-1},A_{\tau_c}^{s-1})
\GETS 
\frac{N\big((X_{\tau_c}^{s-1},x_s),A_{\tau_c}^{s-1}\big) + 1/2}
{\sum_{x'} N\big((X_{\tau_c}^{s-1},x'),A_{\tau_c}^{s-1}\big) + |\X|/2}
\]
\STATE update cost-to-go estimate:\label{algl:jupdate}
\[
\hat{J}(X_{\tau_c}^s,A_{\tau_c}^{s-1})
\GETS
\min_{a_s}\ 
\sum_{x_{s+1}}
\hat{P}\big(x_{s+1}|X_{\tau_c}^s,(A_{\tau_c}^{s-1},a_s)\big)
\left[
g(X_s,a_s,x_{s+1})
+
\alpha
\hat{J}\big((X_{\tau_c}^s,x_{s+1}),(A_{\tau_c}^{s-1},a_s)\big)
\right]
\]
\ENDFOR
\STATE $c\GETS c+1$, $\tau_c \GETS t + 1$
\COMMENT{start the next phrase}
\ENDIF\label{algl:backend}
\ENDFOR
\end{algorithmic}
\end{algorithm*}

At any point in time $t$, if the current phrase started at time $\tau_c$, the
sequence $(X_{\tau_c}^t,A_{\tau_c}^{t-1})$ defines a context which is used to
estimate transition probabilities and cost-to-go function values. To be
precise, given a sequence of observations and actions $(x^\ell,a^{\ell-1})$,
we say the context at time $t$ is $(x^\ell,a^{\ell-1})$ if
$(X_{\tau_c}^t,A_{\tau_c}^{t-1})=(x^\ell,a^{\ell-1})$. For each
$x_{\ell+1}\in\X$ and $a_\ell\in\A$, the algorithm maintains an estimate
$\hat{P}(x_{\ell+1}|x^\ell,a^\ell)$ of the probability of observing
$X_{t+1}=x_{\ell+1}$ at the next time step, given the choice of action
$A_t=a_\ell$ and the current context
$(X_{\tau_c}^t,A_{\tau_c}^{t-1})=(x^\ell,a^{\ell-1})$.  This transition
probability is initialized to be uniform, and subsequently updated using an
empirical estimator based on counts for various realizations of $X_{t+1}$ at
prior visits to the context in question. If $N(x^{\ell+1},a^\ell)$ is the
number of times the context $(x^{\ell+1},a^\ell)$ has been visited prior to
time $t$, then the estimate
\begin{equation}\label{eq:ntop}
  \hat{P}(x_{\ell+1}|x^\ell,a^\ell) 
  = \frac{N(x^{\ell+1},a^\ell) + 1/2}
  {\sum_{x'} N\big((x^{\ell},x'),a^\ell\big) + |\X|/2}
\end{equation}
is used.  This empirical estimator is akin to the update of a Dirichlet-$1/2$
prior with a multinomial likelihood and is similar to that considered by
Krichevsky and Trofimov \cite{Krichevsky81}.

Similarly, at each point in time $t$, given the context
$(X_{\tau_c}^t,A_{\tau_c}^{t-1})=(x^\ell,a^{\ell-1})\in\X^\ell\times\A^{\ell-1}$,
for each $x_{\ell+1}\in\X$ and $a_\ell\in\A$, the quantity
$\hat{J}(x^{\ell+1},a^\ell)$ is an estimate of the cost-to-go if the action
$A_t=a_\ell$ is selected and then observation $X_{t+1}=x_{\ell+1}$ is
subsequently realized. This estimate is initialized to be zero, and
subsequently refined by iterating the dynamic programming operator from
\eqref{eq:bellman} backwards over outcomes that have been previously realized
in the system trajectory, using $\hat{P}$ to estimate the probability of each
possible outcome (line~\ref{algl:jupdate}).

At each time $t$, an action $A_t$ is selected either with the intent to
explore or to exploit.  In the former case, the action is selected uniformly
at random from among all the possibilities (line~\ref{algl:explore}). This
allows the action space to be fully explored and will prove critical in
ensuring the quality of the estimates $\hat{P}$ and $\hat{J}$.  In the latter
case, the impact of each possible action on all future costs is estimated
using the transition probability estimates $\hat{P}$ and the cost-to-go
estimates $\hat{J}$, and the minimizing action is taken acting greedily with
respect to $\hat{P}$ and $\hat{J}$ (line~\ref{algl:exploit}).  A sequence
$\{\gamma_t\}$ controls the relative frequency of actions taken to explore
versus exploit; over time, as the system becomes well-understood, actions are
increasingly chosen to exploit rather than explore.

Note that the active LZ algorithm can be implemented easily using a
tree-like data structure. Nodes at depth $\ell$ correspond to contexts of the
form $(x^\ell,a^{\ell-1})$ that have already been visited. Each such node can
link to at most $|\X||\A|$ child nodes of the form $(x^{\ell+1},a^\ell)$ at
depth $\ell+1$. Each node $(x^{\ell+1},a^\ell)$ maintains a count
$N(x^{\ell+1},a^\ell)$ of how many times it has been seen as a context and
maintains a cost-to-go estimate $\hat{J}(x^{\ell+1},a^\ell)$. The probability
estimates $\hat{P}$ need not be separately stored, since they are readily
constructed from the context counts $N$ according to \eqref{eq:ntop}. Each
phrase interval amounts to traversing a path from the root to a leaf, and
adding an additional leaf. After each such path is traversed, the algorithm
moves backwards along the path
(lines~\ref{algl:backstart}--\ref{algl:backend}) and updates only the counts
and cost-to-go estimates along that path. Note that such an implementation has
linear complexity, and requires a bounded amount of computation and storage
per unit time (or symbol).  

We will shortly establish that the active LZ algorithm achieves the optimal
long-term average cost. Before launching into our analysis, however, we next
consider employing the active LZ algorithm in the context of our running
example of the game of Rock-Paper-Scissors. We have already seen how a player
in this game can minimize his long-term average cost if he knows the
opponent's finite-memory strategy. Armed with the active LZ algorithm, we
can now accomplish the same task \emph{without} knowledge of the opponent's
strategy. In particular, as long as the opponent plays using \emph{some}
finite-memory strategy, the active LZ algorithm will achieve the same
long-term average cost as an optimal response to this strategy.

\begin{ex:rps}[Rock-Paper-Scissors] The active LZ algorithm begins with a
  simple model of the opponent---it assumes that the opponent selects actions
  uniformly at random in every time step, as per line~\ref{algl:pinit}. The
  algorithm thus does not factor in play in future time steps in making
  decisions initially, as per line~\ref{algl:jinit}. As the algorithm
  proceeds, it refines its estimates of the opponent's behavior. For game
  $t+1$, the current context $(X^t_{\tau_c},A^{t-1}_{\tau_c})$ specifies a
  recent history of the game. Given this recent history, algorithm can make a
  prediction of the opponent's next play according to $\hat{P}$, and an
  estimate of the cost-to-go according to $\hat{J}$. These estimates are
  refined as play proceeds and more opponent behavior is observed. If these
  estimates converge to their corresponding true values, the algorithm makes
  decisions (line~\ref{algl:exploit}) that correspond to the optimal decisions
  that would be made if the true transition kernel and cost-to-go function
  were known, as in \eqref{eq:optact}--\eqref{eq:optact2}.
\end{ex:rps}

\subsection{Numerical Experiments with Rock-Paper-Scissors}

Before proceeding with our analysis that establishes the average cost
optimality of the active LZ algorithm, we demonstrate its performance on a
simple numerical example of the Rock-Paper-Scissors game. The example will
highlight the importance of making decisions that optimize long-term costs. 

Consider a simple opponent that plays as follows. If, in the previous
game, the opponent played rock against scissors, the opponent will
play rock again deterministically. Otherwise, the opponent will pick a
play uniformly at random. It is easy to see that an optimal strategy
against such an opponent is to consistently play scissors until
$(\text{rock},\text{scissors})$ occurs, play paper for one game, and
then repeat. Such a strategy incurs an optimal average cost of $-0.25$.

We will compare the performance of the active LZ algorithm against this
opponent versus the performance of an algorithm (which we call `predictive
LZ') based on the Lempel-Ziv predictor of Martinian \cite{Martinian00}. Here,
we use the Lempel-Ziv algorithm to predict the opponent's most likely next
play based on his history, and play the best response. Since Lempel-Ziv offers
both strong theoretical guarantees and impressive practical performance for
the closely related problems of compression and prediction, we would expect
this algorithm would be effective at detecting and exploiting non-random
behavior of the opponent. Note, however, such an algorithm is myopic in that
it is always optimizing one-step costs and does not factor in the effect of
its actions on the opponent's future play.

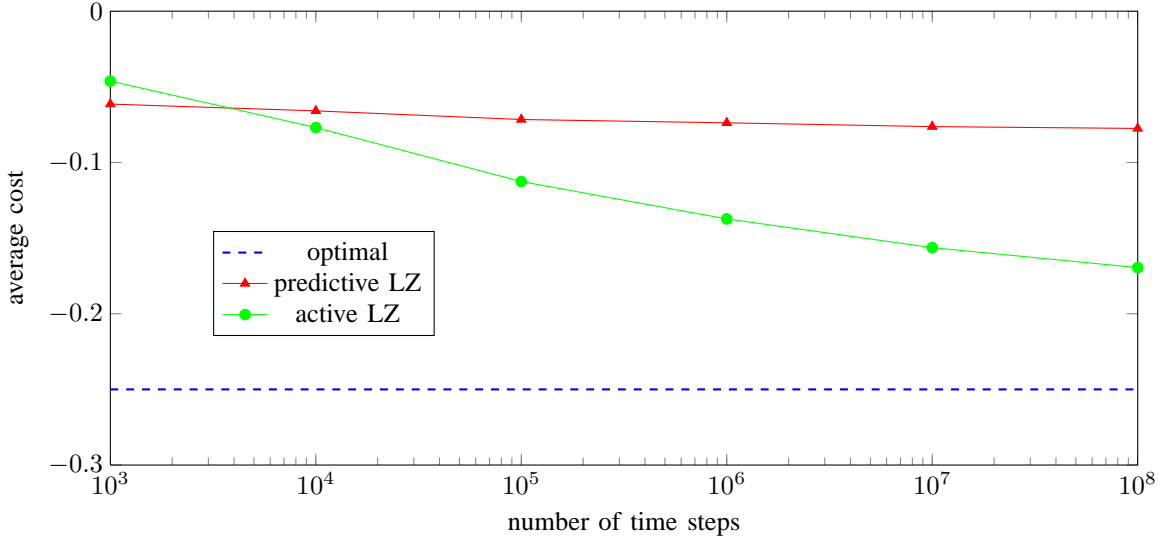
\begin{figure*}[ht]
\begin{center}
\begin{tikzpicture}
  \begin{semilogxaxis}[xlabel=number of time steps,
    ylabel=average cost,
    ymin=-0.3,ymax=0,
    xmin=1000,xmax=100000000,
    width=6in,height=3in,
    legend style={anchor=west, at={(0.1,0.4)}},
    ]
    
    \addplot[thick,dashed,mark=none,blue] coordinates {
      (1000,-0.25)	
      (10000,-0.25)
      (100000,-0.25)	
      (1000000,-0.25)	
      (10000000,-0.25)
      (100000000,-0.25)
    };
    \addlegendentry{optimal}

    \addplot[red,mark=triangle*] coordinates {
      (1000,-0.0614)
      (10000,-0.0659)
      (100000,-0.0716)
      (1000000,-0.0738)
      (10000000,-0.0763)
      (100000000,-0.0775)
    };
    \addlegendentry{predictive LZ}

    \addplot[green,mark=*] coordinates {
      (1000,-0.0462)
      (10000,-0.0769)
      (100000,-0.1126)
      (1000000,-0.1373)
      (10000000,-0.1563)
      (100000000,-0.1695)
    };
    \addlegendentry{active LZ}
  \end{semilogxaxis}
\end{tikzpicture}
\end{center}
\caption{Performance of the active LZ algorithm on Rock-Paper-Scissors
  relative to the predictive LZ algorithm and the optimal
  policy.\label{fi:rps}}
\end{figure*}

In Figure~\ref{fi:rps}, we can see the relative performance of the two
algorithms. The predictive LZ algorithm is able to make some modest
improvements but gets stuck at a fixed level of performance that is
well below optimum. The active LZ algorithm, on the other hand is
able to make consistent improvements. The time required for
convergence to the optimal cost does, however, appear to be
substantial.

\section{Analysis}\label{sec:analysis}

We now proceed to analyze the active LZ algorithm. In particular,
our main theorem, Theorem~\ref{th:opt}, will show that the average
cost incurred upon employing the active LZ algorithm will equal
the optimal average cost, starting at any state.

\subsection{Preliminaries}

We begin with some notation. Recall that, for each $c \geq 1$, $\tau_c$ is the
starting time of the $c$th phrase, with $\tau_1 = 1$. Define $c(t)$ to be
index of the current phrase at time $t$, so that
\[
c(t) \defeq \sup\ \{ c\geq 1\ :\ \tau_c \leq t \}.
\]
At time $t$, the current context will be
$(X^t_{\tau_{c(t)}},A^{t-1}_{\tau_{c(t)}})$. We define the length of the
context at time $t$ to be $d(t) \defeq t - \tau_{c(t)} + 1$.

The active LZ algorithm maintains context counts $N$, probability
estimates $\hat{P}$, and cost-to-go estimates $\hat{J}$. All of these evolve
over time. In order to highlight this dependence, we denote by $N_t$,
$\hat{P}_t$, and $\hat{J}_t$, respectively, the context counts,
probability estimates, and cost-to-go function estimates at time $t$.

Given two probability distributions
$p$ and $q$ over $\X$, define $\TV(p,q)$ to be the total variation distance
\[
\TV(p,q) \defeq \tfrac{1}{2} \sum_x \left| p(x) - q(x) \right|.
\]

\subsection{A Dynamic Programming Lemma}

Our analysis rests on a dynamic programming lemma.  This lemma provides
conditions on the accuracy of the probability estimates $\hat{P}_t$ at time
$t$ that, if satisfied, guarantee that actions generated by acting greedily
with respect to $\hat{P}_t$ and $\hat{J}_t$ are optimal.  It relies heavily on
the fact that the optimal cost-to-go function can be computed by a value
iteration procedure that is very similar to the update for $\hat{J}_t$
employed in the active LZ algorithm.
 
\begin{lemma}\label{le:epsclose}
  Under the active LZ algorithm, there exist constants $\bar{K}\geq 1$ and
  $\bar{\epsilon} \in (0,1)$ so that the following holds: Suppose that, at
  any time $t\geq K$, when the current context is
  $(X^t_{\tau_{c(t)}},A^{t-1}_{\tau_{c(t)}})=(x^s,a^{s-1})$, we have
  \begin{enumerate}
  \item[(i)] 
      The length $s=d(t)$ of the current context is at least $K$.
  \item[(ii)] For all $\ell$ with $s \leq \ell \leq s + \bar{K}$ and all
    $(x_{s+1}^\ell,a_{s}^{\ell-1})$, the context $(x^{\ell},a^{\ell-1})$ has
    been visited at least once prior to time $t$.
  \item [(iii)] For all $\ell$ with $s \leq \ell \leq s + \bar{K}$ and
    all $(x_{s+1}^\ell,a_{s}^{\ell})$, the distribution
    $\hat{P}_t(\cdot|x^\ell,a^\ell)$ satisfies
    \[
    \TV\left(
      \hat{P}_t(\cdot|x^\ell,a^\ell),
      P(\cdot|x^\ell_{\ell-K+1},a^\ell_{\ell-K+1})
    \right)
    \leq
    \bar{\epsilon}.
    \]
  \end{enumerate}
  Then, the action selected by acting greedily with respect to $\hat{P}_t$ and
  $\hat{J}_t$ at time $t$ (as in line~\ref{algl:exploit} of
  the active LZ algorithm) is $\alpha$-discounted optimal. That is, such
  an action is contained in the set of actions
  $\Ascr^*_\alpha(X^t_{t-K+1},A^{t-1}_{t-K+1})$.
\end{lemma}
\begin{IEEEproof}
First, note that there exists a constant $\epsilon > 0$ so that if
$\tilde{P}: \X^K\times\A^K\rightarrow [0,1]$ and $\tilde{J}:
\X^K\times\A^{K-1} \rightarrow \R$ are two arbitrary functions with
\begin{equation}\label{eq:Pclose}
\| \tilde{P}(\cdot|x^K,a^K) - P(\cdot|x^K,a^K) \|_1
< \epsilon,\quad\forall\ x^K,\ a^K,
\end{equation}
\begin{equation}\label{eq:Jclose}
| \tilde{J}(x^K,a^{K-1}) - 
J_\alpha^*(x^K,a^{K-1})
|
< \epsilon,\quad\forall\ x^K,\ a^{K-1},
\end{equation}
then acting greedily with respect to $(\tilde{P},\tilde{J})$ results in
actions that are also optimal with respect to $(P,J_\alpha^*)$---that is, an
optimal policy. The existence of such an $\epsilon$ follows from the
finiteness of the observation and action spaces.

Now, suppose that, at time $t$, 
the hypotheses of the lemma hold for some
$(\bar{\epsilon},\bar{K})$, and that the current context is $(x^s,a^{s-1})$, with
$s=d(t)$. If we can demonstrate that, for every $a_s\in\Ascr$,
\begin{equation}\label{eq:opt1}
\sum_{x_{s+1}}
\left|
\hat{P}_t(x_{s+1}|x^{s},a^{s})
-
P(x_{s+1}|x^{s}_{s-K+1},a^{s}_{s-K+1})
\right| < \epsilon,
\end{equation}
and
\begin{equation}\label{eq:opt2}
\max_{x_{s+1},a_s}\ 
\left|
\hat{J}_t(x^{s+1},a^s)
-
J_\alpha^*(x^{s+1}_{s-K+2},a^{s}_{s-K+2})
\right|
< \epsilon,
\end{equation}
then, by the discussion above, the conclusion of the lemma
holds. \eqref{eq:opt1} is immediate from our hypotheses if $\bar{\epsilon} <
\epsilon/2$.

It remains to establish \eqref{eq:opt2}. In order to do so, fix a choice of
$x_{s+1}$ and $a_s$. To simplify notation in what follows, we will
suppress the dependence of certain probabilities, costs, and value functions
on $(x^{s+1},a^{s})$. In particular, for all $x_{s+2}$ and $a_{s+1}$, define
\[
\hat{P}_t(x_{s+2}|a_{s+1})
\defeq
\hat{P}_t(x_{s+2}|x^{s+1},a^{s+1}),
\]
\[
P(x_{s+2}|a_{s+1})
\defeq
P(x_{s+2}|x^{s+1}_{s-K+2},a^{s+1}_{s-K+2}).
\]
These are, respectively, estimated and true transition probabilities.
Define
\[
g_t(a_{s+1},x_{s+2}) \defeq g(x_s,a_{s+1},x_{s+2})
\]
to be the current cost, and define the value functions
\[
\hat{J}_t(x_{s+2},a_{s+1})
\defeq
\hat{J}_t(x^{s+2},a^{s+1}),
\]
\[
J_\alpha^*(x_{s+2},a_{s+1})
\defeq
J_\alpha^*(x^{s+2}_{s-K+3},a^{s+1}_{s-K+3}).
\]

Then, using the fact that $J_\alpha^*$ solves the Bellman equation
\eqref{eq:bellman} and the recursive definition of $\hat{J}_t$
(line~\ref{algl:jupdate} in the active LZ algorithm), we have
\[
\begin{split}
\lefteqn{
|\hat{J}_t(x^{s+1},a^s) - J_\alpha^*(x^{s+1}_{s-K+2},a^s_{s-K+2}) |
}
\\
& =
\Biggl|
\min_{a_{s+1}}\ 
\sum_{x_{s+2}}
\hat{P}_t(x_{s+2}|a_{s+1})
\\
& 
\qquad\qquad\quad\times
\Big[
g_t(a_{s+1},x_{s+2})
+
\alpha
\hat{J}_t(x_{s+2},a_{s+1})
\Big]
\\
&
\quad\ 
-
\min_{a_{s+1}}\ 
\sum_{x_{s+2}}
P(x_{s+2}|a_{s+1})
\\
&
\qquad\qquad\quad\times
\Big[
g_t(a_{s+1},x_{s+2})
+
\alpha
J_\alpha^*(x_{s+2},a_{s+1})
\Big]
\Biggr|.
\end{split}
\]
Observe that, for any $v,w:\ \Ascr\rightarrow\R$, 
\[
\left| \min_a\ v(a) - \min_a\ w(a) \right| \leq \max_a\ |v(a) - w(a)|.
\]
Then,
\[
\begin{split}
\lefteqn{
|\hat{J}_t(x^{s+1},a^s) - J_\alpha^*(x^{s+1}_{s-K+2},a^s_{s-K+2}) |
}
\\
& \leq
\max_{a_{s+1}}\ 
\Biggl|
\sum_{x_{s+2}}
\hat{P}_t(x_{s+2}|a_{s+1})
\\
&
\qquad\qquad\quad\times
\left[
g_t(a_{s+1},x_{s+2})
+
\alpha
\hat{J}_t(x_{s+2},a_{s+1})
\right]
\\
&
\quad\quad\quad\quad
-
\sum_{x_{s+2}}
P(x_{s+2}|a_{s+1})
\\
&
\qquad\qquad\quad\times
\left[
g_t(a_{s+1},x_{s+2})
+
\alpha
J_\alpha^*(x_{s+2},a_{s+1})
\right]
\Biggr|,
\end{split}
\]

It follows that
\[
\begin{split}
\lefteqn{
|\hat{J}_t(x^{s+1},a^s) - J_\alpha^*(x^{s+1}_{s-K+2},a^s_{s-K+2}) |
}
\\
& \leq
2 g_{\max} \bar{\epsilon} 
\\
& 
\quad
+ 
\alpha
\max_{a_{s+1}}\ 
\Biggl|
\sum_{x_{s+2}}
\Big[
\hat{P}_t(x_{s+2}|a_{s+1})
\hat{J}_t(x_{s+2},a_{s+1})
\\
&
\qquad\qquad\qquad\qquad
-
P(x_{s+2}|a_{s+1})
J_\alpha^*(x_{s+2},a_{s+1})
\Big]
\Biggr|
\\
& \leq
2 g_{\max} \bar{\epsilon} 
\\
&
\quad
+ 
\alpha
\max_{a_{s+1}}\ 
\Biggl|
\sum_{x_{s+2}}
\hat{J}_t(x_{s+2},a_{s+1})
\\
&
\qquad\qquad\qquad\times
\Big[
\hat{P}_t(x_{s+2}|a_{s+1})
-
P(x_{s+2}|a_{s+1})
\Big]
\Biggr|
\\
&
\quad\quad\quad\quad
+
\Biggl|
\sum_{x_{s+2}}
P(x_{s+2}|a_{s+1})
\\
&
\qquad\qquad\qquad\times
\Big[J_\alpha^*(x_{s+2},a_{s+1})
-
\hat{J}_t(x_{s+2},a_{s+1})
\Big]
\Biggr|
\\
\end{split}
\]
Using the fact that $|\hat{J}_t| < g_{\max}/(1-\alpha)$, since it represents a
discounted sum,
\[
\begin{split}
\lefteqn{
|\hat{J}_t(x^{s+1},a^s) - J_\alpha^*(x^{s+1}_{s-K+2},a^s_{s-K+2}) |
}
\\
& \leq
2 g_{\max} \bar{\epsilon}\left( 1 + \frac{\alpha}{1-\alpha}\right) 
\\
&
\quad
+ 
\alpha
\max_{a_{s+1},x_{s+2}}\ 
\left|
J_\alpha^*(x_{s+2},a_{s+1})
-
\hat{J}_t(x_{s+2},a_{s+1})
\right|.
\end{split}
\]
We can repeat this same
analysis on the $|J_\alpha^*(x_{s+2},a_{s+1}) -
\hat{J}_t(x_{s+2},a_{s+1})|$ term. Continuing this $\bar{K}$ times,
we reach the expression
\begin{equation}\label{eq:jrecurse}
\begin{split}
\lefteqn{|\hat{J}_t(x^{s+1},a^s) - J_\alpha^*(x^{s+1}_{s-K+2},a^s_{s-K+2}) |}
\\
& \leq
2 g_{\max} \bar{\epsilon}\left( 1 + \frac{\alpha}{1-\alpha}\right) 
\sum_{\ell=0}^{\bar{K}-1} \alpha^\ell
+ \frac{\alpha^{\bar{K}}g_{\max}}{1-\alpha}
\\
& \leq
\frac{2 g_{\max} \bar{\epsilon}}{1 - \alpha}
\left( 1 + \frac{\alpha}{1-\alpha}\right) 
+ \frac{\alpha^{\bar{K}}g_{\max}}{1-\alpha}.
\end{split}
\end{equation}
It is clear that we can pick $\bar{\epsilon}$ sufficiently small and $\bar{K}$
sufficiently large so that $\bar{\epsilon} < \epsilon/2$ and the right hand
size of \eqref{eq:jrecurse} is less than $\epsilon$.
Such a choice will ensure that \eqref{eq:opt1}--\eqref{eq:opt2} hold, and
hence the requirements of the lemma.
\end{IEEEproof}

Lemma~\ref{le:epsclose} provides sufficient conditions to guarantee when
the active LZ algorithm can be expected to select the correct action given
a current context of $(x^{s},a^{s-1})$. The sufficient conditions are a
requirement the length of the current context, and on the context counts and
probability estimates over all contexts (up to a certain length) that have
$(x^{s},a^{s-1})$ as a prefix.

We would like to characterize when these conditions hold. Motivated by
Lemma~\ref{le:epsclose}, we define the following events for ease of
exposition:

\begin{definition}[$\bar{\epsilon}$-One-Step Inaccuracy]
Define $\Iscr_t^{\bar{\epsilon}}$ to be the event that, at time $t$, at least one of the following holds:
    \begin{enumerate}
\item[(i)] 
$
\TV\left(
\hat{P}_t(\cdot|X^t_{\tau_{c(t)}} ,A^t_{\tau_{c(t)}}),  
P(\cdot|X^t_{t-K+1},A^t_{t-K+1})
\right) > \bar{\epsilon}.
$
\item[(ii)] The current context $(X^t_{\tau_{c(t)}} ,A^{t-1}_{\tau_{c(t)}})$
  has never been visited prior to time $t$.
    \end{enumerate}
\end{definition}

If the event $\Iscr_t^{\bar{\epsilon}}$ holds, then at time $t$ the algorithm
either possesses an estimate of the next-step transition probability
$\hat{P}_t(\cdot|X^t_{\tau_{c(t)}} ,A^t_{\tau_{c(t)}})$ that is more than
$\bar{\epsilon}$ inaccurate relative to the true transition probabilities,
under the total variation metric, or else these probabilities have never been
updated from their initial values.

\begin{definition}[$\bar{\epsilon},\bar{K}$-Inaccuracy]
  Define $\Bscr_t^{\bar{\epsilon},\bar{K}}$ to be the event that, at
  time $t\geq K$, either
  \begin{enumerate}
  \item[(i)] The length $d(t)$ of the current context is less than $K$.
  \item[(ii)] There exist $\ell$ and $(x^\ell,a^\ell)$ such that
    \begin{enumerate}
    \item[(a)] $d(t) \leq \ell \leq d(t) + \bar{K}$.
    \item[(b)] $(x^\ell,a^\ell)$ contains the current context
      $(X^t_{\tau_{c(t)}},A^{t-1}_{\tau_{c(t)}})$ as a prefix, that is,
      \[
      x^{d(t)} = X^t_{\tau_{c(t)}},
      \quad
      a^{d(t)-1} = A^{t-1}_{\tau_{c(t)}}.
      \]
    \item[(c)] The estimated transition probabilities
      $\hat{P}_t(\cdot|x^\ell,a^\ell)$ are more than $\bar{\epsilon}$
      inaccurate, under the total variation metric, and/or the context
      $(x^\ell,a^{\ell-1})$ has never been visited prior to time $t$.
    \end{enumerate}
  \end{enumerate}
\end{definition}

From Lemma~\ref{le:epsclose}, it follows that if the event
$\Bscr_t^{\bar{\epsilon},\bar{K}}$ \emph{does not} hold, then the algorithm
has sufficiently accurate probability estimates in order to make an optimal
decision at time $t$.

Our analysis of the active LZ algorithm proceeds in two broad steps:
\begin{enumerate}
\item In Section~\ref{se:trans}, we establish that $\bar{\epsilon}$-one-step
  inaccuracy occurs a vanishing fraction of the time. Next, we show that this,
  in fact, suffices to establish that $\bar{\epsilon},\bar{K}$-inaccuracy also
  occurs a vanishing fraction of the time. By Lemma~\ref{le:epsclose}, this
  implies that, when the algorithm chooses to exploit, the selected action is
  sub-optimal only a vanishing fraction of the time.
\item In Section~\ref{se:opt}, by further
  controlling the exploration rate appropriately, we can use these results to
  conclude that the algorithm attains the optimal average cost.
\end{enumerate}

\subsection{Approximating Transition Probabilities}
\label{se:trans}

We digress briefly, to discuss a result from the theory of universal
prediction: given an arbitrary sequence $\{y_t\}$, with $y_t \in \Y$ for some
finite alphabet $\Y$, consider the problem of making sequential probability
assignments $Q_{t-1}(\cdot)$ over $\Y$, given the entire sequence observed up to
and including time $t-1$, $y^{t-1}$, so as to minimize the cost function
$\sum_{t=1}^T -\log Q_{t-1}(y_t)$, for some horizon $T$.  It has been shown
by Krichevsky and Trofimov \cite{Krichevsky81} that the assignment
\begin{equation}\label{eq:seqpr}
Q_t(y) \defeq \frac{N_{t}(y) + 1/2}{t + |\Y|/2},
\end{equation}
where $N_t(y)$ is the number of occurrences of the symbol $y$ up to
time $t$, achieves:
\begin{lemma}\label{le:singleletter}
\begin{multline*}
-\sum_{t=1}^T \log Q_{t-1}(y_{t})
-
\min_{q\in\Mscr(\Y)}\ 
\left[
-\sum_{t=1}^T \log q(y_{t})
\right]
\\
\leq 
\frac{|\Y|}{2}\log T + O(1),
\end{multline*}
where the minimization in taken over the set $\Mscr(\Y)$ of all probability
distributions on $\Y$.
\end{lemma}

Lemma~\ref{le:singleletter} provides a bound on the performance of the
sequential probability assignment \eqref{eq:seqpr} versus the
performance of the best constant probability assignment, made with
knowledge of the full sequence $y^T$. Notice that \eqref{eq:seqpr} is
precisely the one-step transition probability estimate employed at
each context by the active LZ algorithm (line~\ref{algl:pupdate}). 

Returning to our original setting, define $p_{\min}$ to be the smallest element of the set of non-zero transition probabilities
\[
\left\{
P(x_{K+1}|x^K,a^K)\ :\ P(x_{K+1}|x^K,a^K) > 0 \right\}.
\]
The proof of the following lemma essentially involves invoking
Lemma~\ref{le:singleletter} at each context encountered by the algorithm, the
use of a combinatorial lemma (Ziv's inequality), and the use of the
Azuma-Hoeffding inequality (see, for example, \cite{Dembo98}). Part of the
proof is motivated by results on Lempel-Ziv based prediction obtained by
Feder et al.\ \cite{Feder92}.

\begin{lemma}\label{le:tvrate}
For arbitrary
$\epsilon' > 0$,
\begin{multline*}
\Pr\left(
\frac{1}{T}
\sum_{t=K}^{T}
\Inb{
\Iscr_t^{\bar{\epsilon}}
}
\geq 
\frac{K_1}{2\bar{\epsilon}^2} \frac{\log\log T}{\log T} + \frac{\epsilon'}{2 \bar{\epsilon}^2}
\right)
\\
\leq
\exp
\left(
-
\frac{T \epsilon'^2}
{8 \log^2\left((2 T + |\X|)/ p_{\min}\right) }
\right),
\end{multline*}
where $K_1$ is a constant that depends only on $|\X|$ and $|A|$.
\end{lemma}
\begin{IEEEproof} 
See Appendix~\ref{app:tvrate}.
\end{IEEEproof}

Lemma~\ref{le:tvrate} controls the fraction of the time that
the active LZ algorithm is $\bar{\epsilon}$-one-step inaccurate. In
particular, Lemma~\ref{le:tvrate} is sufficient to establish that this
fraction of time goes to $0$ (via a use of the first Borel-Cantelli lemma) and
also gives us a rate of convergence. 

It turns out that if the exploration rate $\gamma_t$ decays sufficiently
slowly, this suffices to ensure that the fraction of time the algorithm is
$\bar{\epsilon},\bar{K}$-inaccurate goes to $0$ as well.  To see this, suppose
that the current context at time $t$ is
$(X^t_{\tau_{c(t)}},A^{t-1}_{\tau_{c(t)}})=(x^s,a^{s-1})$, and that the
algorithm is $\bar{\epsilon},\bar{K}$-inaccurate (i.e., the event
$\Bscr^{\bar{\epsilon},\bar{K}}_t$ holds). Then, one of two things must be the
case:
\begin{itemize}
\item The current context length $s$ is less than $K$. We
  will demonstrate that this happens only a vanishing fraction of the time.

\item There exists $(x^\ell,a^\ell)$, with $s \leq \ell
  \leq s + \bar{K}$, so that either the estimated transition probability
  distribution $\hat{P}_t(\cdot|x^\ell,a^\ell)$ is $\bar{\epsilon}$ inaccurate
  under the total variation metric, or the context $(x^\ell,a^{\ell-1})$ has
  never been visited in the past. The probability that the realized sequence
  of future observations and actions $(X^{t+\ell-s}_{t+1},A^{t+\ell-s}_{t})$
  will indeed correspond to $(x^\ell_{s+1},a^\ell_{s})$ is at least
  \[
  p_{\min}^{\ell-s} \prod_{m=t}^{t+\ell-s} \gamma_m,
  \]
  where $p_{\min}$ is the smallest non-zero transition probability. Thus, with
  this minimum probability, a $\bar{\epsilon}$-one-step inaccurate time will
  occur before the time $t+\bar{K}$. Then, if the exploration probabilities
  $\{\gamma_m\}$ decays sufficiently slowly, would be impossible for the
  fraction of $\bar{\epsilon}$-one-step inaccurate times to go to $0$ without
  the fraction of $\bar{\epsilon},\bar{K}$-inaccurate times also going to $0$.
\end{itemize}
By making these arguments precise we can prove the following lemma. The lemma
states that the fraction of time we are at a context wherein the assumptions
of Lemma~\ref{le:epsclose} are not satisfied goes to $0$ almost surely.

\begin{lemma}\label{le:bad}
Assume that
\[
\gamma_t \geq (a_1/\log t)^{1/(a_2\bar{K})},
\]
for arbitrary constants
$a_1 > 0$ and $a_2 > 1$. Further assume that $\{\gamma_t\}$ is non-increasing.
Then,
\[
\lim_{T\tends\infty}
\frac{1}{T}
\sum_{t=K}^{T}
\Inb{
\Bscr^{\bar{\epsilon},\bar{K}}_t
}
= 0,\quad\text{a.s.}
\]
\end{lemma}
\begin{IEEEproof}
  First, we consider the instances of time where the current context length is
  less than $K$. Note that
  \[
  \begin{split}
    \sum_{t=K}^T \I{d(t) < K}  
    & \leq 
    \sum_{c=1}^{c(T)} \sum_{t=\tau_c}^{\tau_{c+1}-1} 
    \I{t - \tau_c + 1 < K}  
    \\
    & 
    \leq
    \sum_{c=1}^{c(T)} K
    = K c(T).
  \end{split}
  \]
  Applying Ziv's inequality (Lemma~\ref{le:ziv}), 
  \begin{equation}\label{eq:nocontext}
    \lim_{T\rightarrow\infty} 
    \frac{1}{T}
    \sum_{t=K}^T \I{d(t) < K}  
    \leq
    \lim_{T\rightarrow\infty} 
    \frac{K C_2}{\log T} = 0.
  \end{equation}

  Next, define $B_t$ to be the event that an $\bar{\epsilon}$-one-step
  inaccurate time occurs between $t$ and $t+\bar{K}$ inclusive, that is
  \[
  B_t \defeq \bigcup_{s = t}^{t+\bar{K}}
  \Iscr_{s}^{\bar{\epsilon}}.
  \]
  It is easy to see that
  \[
  \begin{split}
  \frac{1}{T}
  \sum_{t=K}^{T}
  \Inb{B_{t}}
  &
  \leq 
  \frac{\bar{K}+1}{T}
  \sum_{t=K}^{T+\bar{K}}
  \Inb{
    \Iscr_{t}^{\bar{\epsilon}}
  }
  \\
  &
  \leq
  \frac{\bar{K}+1}{T}
  \sum_{t=K}^{T}
  \Inb{
    \Iscr_{t}^{\bar{\epsilon}}
  }
  +
  \frac{(\bar{K}+1)^2}{T}.
\end{split}
\]
  From Lemma~\ref{le:tvrate}, we immediately have, for arbitrary
  $\epsilon' > 0$,
\begin{equation}
\label{eq:Bt}
\begin{split}
\Pr\Bigg(
\frac{1}{T}
\sum_{t=K}^{T}
\Inb{
B_t
}
& 
\geq 
\frac{(\bar{K}+1) K_1}{2\bar{\epsilon}^2} \frac{\log\log T}{\log T} 
\\
&
\quad
+
\frac{(\bar{K}+1) \epsilon'}{2 \bar{\epsilon}^2}
\\
&
\quad
+
\frac{(\bar{K}+1)^2}{T}
\Bigg) 
\\
& \leq
\exp
\left(
-
\frac{T \epsilon'^2}
{8 \log^2\left((2 T + |\X|)/ p_{\min}\right) }
\right).
\end{split}
\end{equation}

Define $H_t$ to be the event that $\Bscr_t^{\bar{\epsilon},\bar{K}}$
holds, but $d(t) \geq K$. The event $H_t$ holds when, at time $t$, there
exists some context, up to $\bar{K}$ levels below the current context, which
is $\bar{\epsilon}$-one-step inaccurate. Such a context will be visited with
probability at least 
\[
p_{\min}^{\bar{K}} \prod_{m=t}^{t+\bar{K}} \gamma_m \geq 
(p_{\min}\gamma_{t+\bar{K}})^{\bar{K}+1},
\]
in which case $B_t$ holds. Consequently,
\[
\E[\Inb{B_t} | \Fscr_{t}] 
\geq 
(p_{\min}\gamma_{t+\bar{K}})^{\bar{K}+1} \Inb{H_t}.
\]
Since $\gamma_t$ is non-increasing,
\begin{equation}
\label{eq:badbound}
\frac{1}{T} \sum_{t=K}^{T} \E[\Inb{B_t} | \Fscr_t] 
\geq
\frac{
(p_{\min}\gamma_{T+\bar{K}-1})^{\bar{K}+1}
}{T}  
\sum_{t=K}^{T} 
\Inb{
H_t
}
\end{equation}

Now define, for $i=0,1,\dots,\bar{K}-1$ and $n \geq 0$, martingales
$M^{(i)}_n$ adapted to $\Gscr^{(i)}_n = \Fscr_{K+n\bar{K}+i}$, according to
$M^{(i)}_0 = 0$, and, for $n > 0$,
\[
M^{(i)}_n \defeq \sum_{j=0}^{n-1}  \Inb{B_{K+j\bar{K}+i}} - \E[\Inb{B_{K+j\bar{K}+i}} | \Gscr^{(i)}_j].
\]
Since $|M^{(i)}_n - M^{(i)}_{n-1}| \leq 2$, we have via the Azuma-Hoeffding
inequality, for arbitrary $\epsilon'' > 0$,
\begin{equation}
\label{eq:azuma2.1}
\Pr\left(
M^{(i)}_n \geq n\epsilon''
\right)
\leq 
\exp\left(-n {\epsilon''}^2 / 8\right)
\end{equation}

For each $i$, let $n_i(T)$ be the largest integer such that $K + n_i(T)\bar{K}
+ i \leq T$, so that
\[
\sum_{t=K}^{T} \Inb{B_t} - \E[\Inb{B_{t}} | \Fscr_t] 
= \sum_{i=0}^{\bar{K}-1} M^{(i)}_{n_i(T)}.
\]
Since $n_i(T) \leq \frac{T}{\bar{K}}$, the union bound along with
\eqref{eq:azuma2.1} then implies that:
\begin{equation}
\label{eq:azuma2.2}
\begin{split}
\lefteqn{
\Pr\left(
\sum_{t=K}^{T} \Inb{B_t} - \E[\Inb{B_{t}} | \Fscr_t] \geq T\epsilon''
\right)
}
\\
&
\leq 
\sum_{i=0}^{\bar{K}-1} \Pr\left(
M^{(i)}_{n_i(T)} \geq T\epsilon''/\bar{K}
\right)
\\
&
\leq
\sum_{i=0}^{\bar{K}-1} \exp\left(- T^2{\epsilon''}^2 / 8\bar{K}^2n_i(T)\right)
\\
&
\leq
\bar{K}\exp\left(- T{\epsilon''}^2 / 8\bar{K}\right).
\end{split}
\end{equation}

Now, define 
\[
\begin{split}
\kappa(T)
& 
\defeq
\frac{1}{(p_{\min} \gamma_{T+\bar{K}-1})^{\bar{K}+1}}
\Bigg[
\frac{(\bar{K}+1)K_1}{2\bar{\epsilon}^2} \frac{\log\log T}{\log T} 
\\
&
\qquad\qquad
+ \frac{(\bar{K}+1)\epsilon'(T)}{2 \bar{\epsilon}^2}
+ \frac{(\bar{K}+1)^2}{T}
+ \epsilon''(T)
\Bigg],
\end{split}
\]
with
\[
\epsilon'(T)
\defeq
\frac{1}{\log T},
\quad
\epsilon''(T)
\defeq
\frac{1}{\log T}.
\]
It follows from \eqref{eq:Bt}, \eqref{eq:badbound}, and \eqref{eq:azuma2.2} that
\[
\begin{split}
\lefteqn{
\Pr\left(
\frac{1}{T}  
\sum_{t=K}^{T} 
\Inb{
H_t
}
\geq 
\kappa(T)
\right)
}
\\
&
\leq
\exp
\left(
-
\frac{T}
{8 \log^4\left((2 T + |\X|)/ p_{\min}\right) }
\right)
\\
&
\quad
+
\bar{K}\exp\left(-\frac{T}{8\bar{K}\log^2 T}\right).
\end{split}
\]
By the first Borel-Cantelli lemma,
\[
\Pr\left(
\frac{1}{T}  
\sum_{t=K}^{T-1} 
\Inb{
H_t
}
\geq 
\kappa(T),\ \text{i.o.}
\right)
= 
0.
\]
Note that the hypothesis on $\gamma_t$ implies that $\kappa(T) \tends
0$ as $T\tends \infty$. Then,
\begin{equation}\label{eq:Hvanish}
\lim_{T\rightarrow\infty}
\frac{1}{T}  
\sum_{t=K}^{T-1} 
\Inb{
H_t
}
= 
1,
\quad\text{a.s.}
\end{equation}

Finally, note that
\[
\frac{1}{T}
\sum_{t=K}^{T}
\Inb{
\Bscr^{\bar{\epsilon},\bar{K}}_t
}
\leq
\frac{1}{T}
\sum_{t=K}^{T}
\I{d(t)<K}
+
\frac{1}{T}
\sum_{t=K}^{T}
\Inb{
H_t
}.
\]
The result then follows from \eqref{eq:nocontext} and \eqref{eq:Hvanish}.
\end{IEEEproof}

\subsection{Average Cost Optimality}
\label{se:opt}

Observe that if the active LZ algorithm chooses an action that is
non-optimal at time $t$, that is,
\[
 A_t \notin
\Ascr_\alpha^*(X^t_{t-K+1},A^{t-1}_{t-K+1}),
\]
then, either the event
$\Bscr^{\bar{\epsilon},\bar{K}}_t$ holds or the algorithm chose
to explore. Lemma~\ref{le:bad} guarantees that the first possibility
happens a vanishing fraction of time. Further, if $\gamma_t\downarrow
0$, then the algorithm will explore a vanishing fraction of
time. Combining these observations give us the following theorem.

\begin{theorem}\label{th:zerofrac}
Assume that
\[
\gamma_t \geq (a_1/\log t)^{1/(a_2\bar{K})},
\]
for arbitrary constants
$a_1 > 0$ and $a_2 > 1$. Further, assume that $\gamma_t\downarrow 0$.
Then,
\[
\lim_{T \tends \infty} 
\frac{1}{T}
\sum_{t=K}^{T}
\I{A_t \notin
\Ascr_\alpha^*(X^t_{t-K+1},A^{t-1}_{t-K+1})}
=
0,\quad\text{a.s.}
\]
\end{theorem}
\begin{IEEEproof}
  Given a sequence of independent bounded random variables $\{Z_n\}$,
  with $\E[Z_n] \tends 0$, 
  \[
  \lim_{N\tends\infty} \frac{1}{N}\sum_{n=1}^N Z_n = 0, \quad\text{a.s.}
  \]
  This follows, for example, from the Azuma-Hoeffding inequality followed by
  the first Borel-Cantelli lemma.  This immediately yields
  \begin{equation}
    \label{eq:exploration}
    \lim_{T \tends \infty} 
    \frac{1}{T}
    \sum_{t=k}^{T-1}
    \I{\text{exploration at time $t$}}
    \tends
    0,\quad\text{a.s.},
  \end{equation}
  provided $\gamma_t \tends 0$ (note that the choice of exploration at
  each time $t$ is independent of all other events).
  Now observe that 
  \begin{multline*}
  \left\{ A_t \notin
    \Ascr_\alpha^*(X^t_{t-K+1},A^{t-1}_{t-K+1}) \right\}
  \\
  \subset
  \Bscr^{\bar{\epsilon},\bar{K}}_{t-1}
  \cup
  \{ \text{exploration at time $t$}\}.
\end{multline*}
Combining \eqref{eq:exploration} with Lemma~\ref{le:bad}, the result
  follows.
\end{IEEEproof}

Assumption~\ref{as:eqavgcost} guarantees the optimal average cost is
$\lambda^*$, independent of the initial state of the Markov chain, and that
there exists a stationary policy that achieves the optimal average cost
$\lambda^*$. By the ergodicity theorem, under such a optimal policy,
\begin{equation}\label{eq:erg}
\lim_{T\tends\infty} \frac{1}{T} \sum_{t=1}^T g(X_t,A_t,X_{t+1}) = \lambda^*,
\quad\text{a.s.}
\end{equation}
On the other hand, Theorem~\ref{th:zerofrac} suggests that, under
the active LZ algorithm, the fraction of time at which non-optimal
decisions are made vanishes asymptotically.  Combining these facts yields our
main result.

\begin{theorem}\label{th:opt}
Assume that
\[
\gamma_t \geq (a_1/\log t)^{1/(a_2\bar{K})},
\]
for arbitrary constants
$a_1 > 0$ and $a_2 > 1$, and that $\gamma_t\downarrow 0$.
Then, for $\alpha \in (0,1)$ sufficiently close to $1$,
\[
\lim_{T\tends\infty} \frac{1}{T}\sum_{t=1}^T g(X_t,A_t,X_{t+1}) = \lambda^*,
\quad\text{a.s.,}
\]
under the active LZ algorithm.  Hence, the active LZ algorithm
achieves an asymptotically optimal average cost regardless of the
underlying transition kernel.
\end{theorem}
\begin{IEEEproof}
  Without loss of generality, assume that the cost $g(X_t,A_t,X_{t+1})$ does
  not depend on $X_{t+1}$. 

Fix $\epsilon > 0$, and consider an interval of time
$T_\epsilon>K$. For each $(x^K,a^K)\in\X^K\times\A^K$, define a
coupled process $(\tilde{X}_t(x^k,a^k), \tilde{A}_t(x^K,a^K))$ as
follows. For every integer $n$, set
\[
\tilde{X}^{(n-1)T_\epsilon+K}_{(n-1)T_\epsilon+1}(x^K,a^K)=x^K_1,
\]
and
\[
\tilde{A}^{(n-1)T_\epsilon+K}_{(n-1)T_\epsilon+1}(x^K,a^K)=a^K_1.
\]
For all other times $t$, the coupled processes will choose actions
according to an optimal stationary policy, that is
\[
\tilde{A}_t(x^K,a^K) \in 
\Ascr_\alpha^*
\left(\tilde{X}^{t}_{t-K+1}(x^K,a^K),\tilde{A}^{t-1}_{t-K+1}(x^K,a^K)\right).
\]
Without loss of generality, we will assume that the choice of action is
unique.

Now, for each $n$ there will be exactly one $(x^K,a^K)$ that matches the
original process $(X_t,A_t)$ over times $(n-1)T_\epsilon+1 \leq t \leq
(n-1)T_\epsilon+K$, that is,
\[
(x^K,a^K)
=\left(X^{(n-1)T_\epsilon+K}_{(n-1)T_\epsilon+1},A^{(n-1)T_\epsilon+K}_{(n-1)T_\epsilon+1}
\right).
\] 
For the process indexed by $(x^K,a^K)$, for $(n-1)T_\epsilon+K < t \leq
nT_\epsilon$, if
\[
\left(\tilde{X}^{t-1}_{t-K}(x^K,a^K),\tilde{A}^{t-1}_{t-K}(x^K,a^K)\right)
=(X^{t-1}_{t-K},A^{t-1}_{t-K}),
\]
then set $\tilde{X}_t(x^K,a^K)=X_t$. Otherwise, allow $\tilde{X}_t(x^K,a^K)$
to evolve independently according to the process transition
probabilities. Similarly, allow all other the processes to evolve
independently according to the proper transition probabilities.

Define
\[
G_n(x^k,a^k) \defeq
\frac{1}{T_\epsilon} 
\sum_{t=(n-1)T_\epsilon+1}^{nT_\epsilon}
g\left(\tilde{X}_t(x^K,a^K),\tilde{A}_t(x^K,a^K)\right).
\]
Note that each $G_n(x^K,a^K)$ is the average cost under an {\em optimal}
policy.  Therefore, because of \eqref{eq:erg}, we can pick $T_\epsilon$ large
enough so that for any $n$,
\begin{equation}\label{eq:gmax}
\E\left[
\max_{x^K,a^K}\ 
\left|
G_n(x^K,a^K) 
-
\lambda^*
\right|
\right]
< \epsilon.
\end{equation}

Define $\Zscr_n$ to be the event that,
within the $n$th interval, the algorithm chooses a non-optimal
action. That is,
\[
\Zscr_n \defeq \left\{ \exists\ t,\ (n-1)T_\epsilon < t \leq nT_\epsilon,\ 
A_t \notin \Ascr_\alpha^*(X^t,A^{t-1}) \right\}.
\]
Set
\[
E_N = \frac{1}{N} \sum_{n=1}^N \Inb{\Zscr_n}.
\]
Then,
\[
\begin{split}
\lefteqn{
\left|
\frac{1}{NT_\epsilon} \sum_{t=1}^{NT_\epsilon} (g(X_t,A_t) - \lambda^*)
\right|
}
\\
& \leq
\frac{\max(|g_{\max} - \lambda^*|, \lambda^*) E_N }{N}
\\
& \quad
+
\left|
\frac{1}{NT_\epsilon} \sum_{n=1}^N 
(1 - \Inb{\Zscr_n}) \sum_{t=(n-1)T_\epsilon+1}^{nT_\epsilon} (g(X_t,A_t) - \lambda^*)
\right|.
\end{split}
\]
Note that, from Theorem~\ref{th:zerofrac}, $E_N/N\tends 0$ almost
surely as $N \tends \infty$. Thus,
\[
\begin{split}
\lefteqn{
\limsup_{N\tends\infty}
\left|
\frac{1}{NT_\epsilon} \sum_{t=1}^{NT_\epsilon} (g(X_t,A_t) - \lambda^*)
\right|
}
\\
&
\leq
\limsup_{N\tends\infty}
\frac{1}{N} \sum_{n=1}^N 
(1 - \Inb{\Zscr_n}) 
\\
&
\qquad\qquad\qquad\times
\left|
\frac{1}{T_\epsilon}
\sum_{t=(n-1)T_\epsilon+1}^{nT_\epsilon} (g(X_t,A_t) - \lambda^*)
\right|.
\end{split}
\]
Notice that when $\Inb{\Zscr_n}=0$, we have for some $(x^K,a^K)$ that 
$\tilde{X}_t(x^K,a^K)=X_t$ for all  $(n-1)T_\epsilon < t \leq nT_\epsilon$. Thus,
\[
\begin{split}
\lefteqn{
\limsup_{N\tends\infty}
\left|
\frac{1}{NT_\epsilon} \sum_{t=1}^{NT_\epsilon} (g(X_t,A_t) - \lambda^*)
\right|
}
\\
&
\leq
\limsup_{N\tends\infty}
\frac{1}{N} \sum_{n=1}^N 
(1 - \Inb{\Zscr_n}) 
\max_{x^K,a^K}\ 
\left|
G_n(x^K,a^K) 
- \lambda^*
\right|
\\
&
\leq
\limsup_{N\tends\infty}
\frac{1}{N} \sum_{n=1}^N 
\max_{x^K,a^K}\ 
\left|
G_n(x^K,a^K) 
- \lambda^*
\right|
.
\end{split}
\]
However, the variables
\[
\max_{x^K,a^K}\ 
\left|
G_n(x^K,a^K) 
- \lambda^*
\right|
\]
are independent and identically distributed as $n$ varies. Thus, by
the Strong Law of Large Numbers and \eqref{eq:gmax},
\[
\limsup_{T\tends\infty}
\left|
\frac{1}{T} \sum_{t=1}^{T} (g(X_t,A_t) - \lambda^*)
\right|
\leq
\epsilon,
\]
with probability 1.
Since $\epsilon$ was arbitrary, the result follows.
\end{IEEEproof}

\subsection{Choice of Discount Factor}

Given a choice of $\alpha$ sufficiently close to $1$, the optimal
$\alpha$-discounted cost policy coincides with the average cost
optimal policy. Our presentation thus far has assumed knowledge of
such an $\alpha$. For a given $\alpha$, under the assumptions of
Theorem~\ref{th:zerofrac}, The active LZ algorithm is guaranteed
to take $\alpha$-discounted optimal actions a fraction $1$ of the time
which for an ad-hoc choice of $\alpha$ sufficiently close to $1$ is
likely to yield good performance. Nonetheless, one may use a
`doubling-trick' in conjunction with the active LZ algorithm to
attain average cost optimality without knowledge of $\alpha$.  In
particular, consider the following algorithm that uses
the active LZ algorithm, with the choice of $\{ \gamma_t\}$ as
stipulated by Theorem~\ref{th:zerofrac}, as a subroutine:
\begin{algorithm}
\caption{The active LZ with a doubling scheme.}
\label{alg:doubling}
\begin{algorithmic}[1]
\FOR {non-negative integers $k$}
\FOR {each time $2^k \leq t' < 2^{k+1}$}
\STATE Apply the active LZ algorithm (Algorithm~\ref{alg:lzlearn}) with $\alpha = 1- \beta_k$, and time index $t=t'-2^k$.
\ENDFOR
\ENDFOR
\end{algorithmic}
\end{algorithm}

Here $\beta_k$ is a sequence that approaches $0$ sufficiently
slowly. One can show that if $\beta_k =
\Omega(1/\log\log k)$, then the above scheme achieves average cost
optimality. A rigorous proof of this fact would require repetition of
arguments we have used to prove earlier results.  As such, we only
provide a sketch that outlines the steps required to establish average
cost optimality:

We begin by noting that in the $k$th epoch of
Algorithm~\ref{alg:doubling}, one choice (so that Lemma
\ref{le:epsclose} remains true) is to let $\bar{\epsilon}_k,
\bar{K}_k$ grow as $\alpha$ approaches $1$ according to
$\bar{\epsilon}_k= \Omega(1)$ and $\bar{K}_k =\Omega(1/\beta_k)$
respectively. If $\beta_k = \Omega(1/\log\log k)$, then for the $k$th
epoch of Algorithm~\ref{alg:doubling}, Lemma \ref{le:bad} is easily
modified to show that with high probability the greedy action is
suboptimal over less than $2^k \kappa(2^k)$ time steps where
$\kappa(2^k) = O((\log\log 2^k)^3/\log 2^k)$. The Borel-Cantelli Lemma
may then be used to establish that beyond some finite epoch, over all
subsequent epochs $k$, the greedy action is suboptimal over at most
$2^k \kappa(2^k)$ time steps. Provided $\beta_k \tends 0$, this
suffices to show that the greedy action is optimal a fraction $1$ of
the time. Provided one decreases exploration probabilities
sufficiently quickly, this in turn suffices to establish average cost
optimality.

\subsection{On the Rate of Convergence}

We limit our discussion to the rate at which the fraction of time
the active LZ algorithm takes sub-optimal actions goes to zero; even
assuming one selects optimal actions at every point in time, the rate at which
average costs incurred converge to $\lambda^*$ are intimately related to the
structure of $P$ which is a somewhat separate issue.
Now the proofs of Lemma~\ref{le:bad} and Theorem~\ref{th:zerofrac} tell us
that the fraction of time the active LZ algorithm selects sub-optimal
actions goes to zero at a rate that is $O((1/ \log T)^{c})$ where $c$ is some
constant less than $1$. The proofs of Lemmas~\ref{le:tvrate} and \ref{le:bad}
reveal that the determining factor of this rate is effectively the rate at
which the transition probability estimates provided by $\hat{P}$ converge to
their true values. Thus while the rate at which the fraction of sub-optimal
action selections goes to zero is slow, this rate isn't surprising and is
shared with many Lempel-Ziv schemes used in prediction and compression.

A natural direction for further research is to explore the effect of replacing
the LZ-based context tree data structure by the context-tree weighting method
of Willems et al.\ \cite{Willems95}. It seems plausible to expect that such an
approach will yield algorithms with significantly improved convergence rates,
as is the case in data compression and prediction.

\section{Conclusion}\label{sec:future}

We have presented and established the asymptotic optimality of a
Lempel-Ziv inspired algorithm for learning. The algorithm is a natural
combination of ideas from information theory and dynamic
programming. We hope that these ideas, in particular the use of a
Lempel-Ziv tree to model an unknown probability distribution, can find
other uses in reinforcement learning.

One interesting special case to consider is when the next observation
is Markovian given the past $K$ observations and only the latest
action. In this case, a variation of the active LZ algorithm that
uses contexts of the form $(x^s,a)$ could be used. Here, the resulting
tree would have exponentially fewer nodes and would be much quicker to
converge to the optimal policy.

A number of further issues are under consideration. It would be of great
interest to develop theoretical bounds for the rate of convergence.  Also, it
would be natural to extend the analysis of our algorithms to systems with
possibly infinite dependence on history. One such extension would be to mixing
models, such as those considered by Jacquet et al.\ \cite{Jacquet02}.  Another
would be to consider the the optimal control of a partially observable Markov
decision process.

\bibliographystyle{IEEEtran}
\bibliography{IEEEabrv,nips-ul}

\begin{IEEEbiography}{Vivek F. Farias}
  Vivek Farias is the Robert N. Noyce Career Development Assistant Professor
  of Management at MIT. He received a Ph. D. in Electrical Engineering from
  Stanford University in 2007. He is a recipient of an IEEE Region 6
  Undergraduate Student Paper Prize (2002), a Stanford School of Engineering
  Fellowship (2002), an INFORMS MSOM Student Paper Prize (2006) and an MIT
  Solomon Buchsbaum Award (2008).
\end{IEEEbiography}

\begin{IEEEbiography}{Ciamac C. Moallemi}
  Ciamac C. Moallemi is an Assistant Professor at the Graduate School
  of Business of Columbia University, where he has been since 2007. He
  received SB degrees in Electrical Engineering \& Computer Science
  and in Mathematics from the Massachusetts Institute of Technology
  (1996). He studied at the University of Cambridge, where he earned a
  Certificate of Advanced Study in Mathematics, with distinction
  (1997). He received a PhD in Electrical Engineering from Stanford
  University (2007). He is a member of the IEEE and INFORMS. He is the
  recipient of a British Marshall Scholarship (1996) and a Benchmark
  Stanford Graduate Fellowship (2003).
\end{IEEEbiography}

\begin{IEEEbiography}{Benjamin Van Roy}
  Benjamin Van Roy is an Associate Professor of Management Science and
  Engineering, Electrical Engineering, and, by courtesy, Computer
  Science, at Stanford University. He has held visiting positions as
  the Wolfgang and Helga Gaul Visiting Professor at the University of
  Karlsruhe and as the Chin Sophonpanich Foundation Professor of
  Banking and Finance at Chulalongkorn University.  He received the SB
  (1993) in Computer Science and Engineering and the SM (1995) and PhD
  (1998) in Electrical Engineering and Computer Science, all from
  MIT. He is a member of INFORMS and IEEE. He has served on the
  editorial boards of Discrete Event Dynamic Systems, Machine
  Learning, Mathematics of Operations Research, and Operations
  Research.  He has been a recipient of the MIT George C. Newton
  Undergraduate Laboratory Project Award (1993), the MIT Morris
  J. Levin Memorial Master's Thesis Award (1995), the MIT George
  M. Sprowls Doctoral Dissertation Award (1998), the NSF CAREER Award
  (2000), and the Stanford Tau Beta Pi Award for Excellence in
  Undergraduate Teaching (2003). He has been a Frederick E. Terman
  Fellow and a David Morgenthaler II Faculty Scholar.
\end{IEEEbiography}

\begin{IEEEbiography}{Tsachy Weissman}
  Tsachy Weissman obtained his undergraduate and graduate degrees from the
  Department of electrical engineering at the Technion. Following his
  graduation, he has held a faculty position at the Technion, and postdoctoral
  appointments with the Statistics Department at Stanford University and with
  Hewlett-Packard Laboratories. Since the summer of 2003 he has been on the
  faculty of the Department of Electrical Engineering at Stanford. Since the
  summer of 2007 he has also been with the Department of Electrical
  Engineering at the Technion, from which he is currently on leave.

  His research interests span information theory and its applications, and
  statistical signal processing. He is inventor or co-inventor of several
  patents in these areas and involved in a number of high-tech companies as a
  researcher or member of the technical board.

  His recent prizes include the NSF CAREER award, a Horev fellowship for
  leaders in Science and Technology, and the Henry Taub prize for excellence
  in research.  He is a Robert N. Noyce Faculty Scholar of the School of
  Engineering at Stanford, and a recipient of the 2006 IEEE joint IT/COM
  societies best paper award.
\end{IEEEbiography}

\appendices

\section{Proof of Lemma~\ref{le:tvrate}}
\label{app:tvrate}

An important device in the proof of Lemma~\ref{le:tvrate} the following
combinatorial lemma. A proof can be found in Cover and Thomas \cite{Cover91}.
\begin{lemma}[Ziv's Inequality]\label{le:ziv}
The number of contexts seen by time $T$, $c(T)$, satisfies
\[
c(T) \leq \frac{C_2 T}{\log T},
\]
where
$C_2$ is a constant that depends only on $|\X|$ and $|\A|$.
\end{lemma}

Without loss of generality, assume that $X_t$ and $A_t$ take some fixed but
arbitrary values of $-K + 2 \leq t \leq 0$, so that the expression
$P(X_{t+1}|X^{t}_{t-K+1},A^{t}_{t-K+1})$ is well-defined for all $t \geq 1$.
We will use Lemma~\ref{le:singleletter} to show: 
\begin{lemma}\label{le:contextb}
\[
\begin{split}
\lefteqn{
-\sum_{t=1}^{T} \log \hat{P}_t(X_{t+1}|X^{t}_{\tau_{c(t)}},A^{t}_{\tau_{c(t)}})
}
\\
& \leq
-\sum_{t=1}^{T} \log P(X_{t+1}|X^{t}_{t-K+1},A^{t}_{t-K+1})
\\
&
\quad
+
\bar{K}_1 T\frac{ \log\log T}{\log T},
\end{split}
\] 
where $\bar{K}_1$ is a positive constant that depends only on $|\X|$ and $|\A|$.
\end{lemma}
\begin{IEEEproof}
  Observe that the probability assignment made by our algorithm is
  equivalent to using \eqref{eq:seqpr} at every context. In
  particular, at every time $t$,
\begin{multline*}
\hat{P}_t(X_{t+1}|X^{t}_{\tau_{c(t)}},A^{t}_{\tau_{c(t)}}) 
\\
= 
\frac
{N_t(X^{t+1}_{\tau_{c(t)}},A^{t}_{\tau_{c(t)}}) + 1/2}
{\sum_x N_t((X^{t}_{\tau_{c(t)}},x),A^{t}_{\tau_{c(t)}}) + |\X|/2}
\end{multline*}
For each $(x^j,a^j)$, define $\Tscr_T(x^j,a^j)$ to be the set of times
\[
\Tscr_T(x^j,a^j) \defeq \left\{  t\ :\ 1 \leq t \leq T,\ 
(X^{t}_{\tau_{c(t)}},A^{t}_{\tau_{c(t)}})
 = (x^j,a^j)
\right\}.
\]
It follows from Lemma~\ref{le:singleletter} that
\[
\begin{split}
\lefteqn{
-\sum_{t\in\Tscr_T(x^j,a^j)}
\log \hat{P}_t(X_{t+1}|X^{t}_{\tau_{c(t)}},A^{t}_{\tau_{c(t)}}) 
}
\\
& \leq
\min_{p \in \Mscr(\X)}\ 
-
\sum_{t\in\Tscr_T(x^j,a^j)}
 \log p(X_{t+1})
\\
&
\quad
+ \frac{|\X|}{2} \log |\Tscr_T(x^j,a^j)| + C_1.
\end{split}
\]
Summing this expression over all distinct $(x^j,a^j)$ that have
occurred up to time $T$,
\begin{equation}\label{eq:sum1}
\begin{split}
\lefteqn{
-\sum_{t=1}^{T} 
\log \hat{P}_t(X_{t+1}|X^{t}_{\tau_{c(t)}},A^{t}_{\tau_{c(t)}}) 
}
\\
& \leq
\sum_{(x^j,a^j)}
\left[
\min_{p \in \Mscr(\X)}\ 
-
\sum_{t\in\Tscr(x^j,a^j)}
 \log p(X_{t+1})
\right]
\\
&
\quad
+
\sum_{(x^j,a^j)} 
\left[
\frac{|\X|}{2} \log |\Tscr_T(x^j,a^j)| 
+
C_1
\right]
\\
&
\leq
-\sum_{t=1}^{T} 
\log P(X_{t+1}|X^{t}_{t-K+1},A^{t}_{t-K+1})
\\
&
\quad
+
\sum_{(x^j,a^j)} 
\left[
\frac{|\X|}{2} \log |\Tscr_T(x^j,a^j)| 
+
C_1
\right].
\end{split}
\end{equation}

Now, $c(T)$ is the total number of distinct contexts that have occurred up to
time $T$. Note that this is also precisely the number of distinct $(x^j,a^j)$
with $|\Tscr_T(x^j,a^j)| > 0$. Then, by the concavity of  $\log(\cdot)$,
\begin{multline*}
\sum_{(x^j,a^j)} 
\left[
\frac{|\X|}{2} \log |\Tscr_T(x^j,a^j)| 
+C_1
\right]
\\
\leq
\frac{|\X| c(T)}{2} \log\frac{T}{c(T)} + C_1c(T).
\end{multline*}
Applying Lemma~\ref{le:ziv},
\begin{equation}\label{eq:sum2}
\begin{split}
\lefteqn{
\sum_{(x^j,a^j)} 
\left[
\frac{|\X|}{2} \log |\Tscr_T(x^j,a^j)| 
+C_1
\right]
}
\\
&
\leq 
\frac{C_2 |\X|}{2}
\frac{T}{\log T}
\left[ \log \log T - \log C_2 \right]
\\
&
\quad 
+ C_1 C_2 \frac{T}{\log T}.
\end{split}
\end{equation}
The lemma follows by combining \eqref{eq:sum1} and \eqref{eq:sum2}.
\end{IEEEproof}

For the remainder of this section, define $\Delta_t$ to be the
Kullback-Leibler distance between the estimated and true transition
probabilities at time $t$, that is
\[
\Delta_t \defeq
D\left(
P(\cdot|X^{t}_{t-K+1},A^{t}_{t-K+1})
\big\|
\hat{P}_t(\cdot|X^{t}_{\tau_{c(t)}},A^{t}_{\tau_{c(t)}})
\right).
\]

\begin{lemma}
\label{le:contextub}
For arbitrary $\epsilon' > 0$, 
\[
\begin{split}
\Pr\Bigg(
&
\frac{1}{T}
\sum_{t=1}^{T}
\left[
\log 
\frac{\hat{P}_t(X_{t+1}|X^{t}_{\tau_{c(t)}},A^{t}_{\tau_{c(t)}})}
{P(X_{t+1}|X^{t}_{t-K+1},A^{t}_{t-K+1})}
+
\Delta_t
\right]
\\
&
\qquad\geq 
\epsilon'
\Bigg)
\\
& \leq
\exp
\left(
-
\frac{T \epsilon'^2}
{8 \log^2\left((2 T + |\X|)/ p_{\min}\right) }
\right).
\end{split}
\]
\end{lemma}
\begin{IEEEproof}
  Define, for $T \geq 0$, a process $\{M_T\}$ adapted to
  $\Fscr_{T+1}$ as follows: set with $M_0=0$, and, for $T >1$,
\[
\begin{split}
M_T
& \defeq
\sum_{t=1}^{T}
\log \frac{\hat{P}_t(X_{t+1}|X^{t}_{\tau_{c(t)}},A^{t}_{\tau_{c(t)}})}{P(X_{t+1}|X^{t}_{t-K+1},A^{t}_{t-K+1})}
\\
&
\quad
-
\sum_{t=1}^{T}
\E
\left[
\left.
\log \frac{\hat{P}_t(X_{t+1}|X^{t}_{\tau_{c(t)}},A^{t}_{\tau_{c(t)}})}{P(X_{t+1}|X^{t}_{t-K+1},A^{t}_{t-K+1})}
\right|
\Fscr_{t}
\right]
\\
&
=
\sum_{t=1}^{T}
\left(
\log \frac{\hat{P}_t(X_{t+1}|X^{t}_{\tau_{c(t)}},A^{t}_{\tau_{c(t)}})}{P(X_{t+1}|X^{t}_{t-K+1},A^{t}_{t-K+1})}
+ \Delta_t
\right).
\end{split}
\]
It is clear that $M_T$ is a martingale with $\E[M_T] = 0$. 
Further, 
\[
0 \geq
\log \hat{P}_t(X_{t+1} | X^{t}_{\tau_{c(t)}}, A^{t}_{\tau_{c(t)}}) 
\geq \log(1/(2t + |\X|)).
\]
and
\[
0 \geq
\log P(X_{t+1} | X^{t}_{t-K+1}, A^{t}_{t-K+1}) 
\geq \log p_{\min},
\]
so that
\[
|M_T - M_{T-1}| \leq 2 \log\left(\frac{2 T + |\X|}{p_{\min}}\right).
\]
An application of the Azuma-Hoeffding inequality then yields, for arbitrary $\epsilon' >
0$,
\[
\begin{split}
\Pr
\left(
\frac{M_T}{T}
\geq 
\epsilon'
\right)
&
\leq
\exp
\left(
-
\frac{T^2 \epsilon'^2}
{8 \sum_{t=1}^T  \log^2\left((2 T + |\X|)/ p_{\min}\right) }
\right)
\\
&
\leq
\exp
\left(
-
\frac{T \epsilon'^2}
{8 \log^2\left((2 T + |\X|)/ p_{\min}\right) }
\right).
\end{split}
\]
\end{IEEEproof}

We are now ready to prove Lemma~\ref{le:tvrate}.
\begin{le:tvrate}
For arbitrary
$\epsilon' > 0$,
\begin{multline*}
\Pr\left(
\frac{1}{T}
\sum_{t=K}^{T}
\Inb{
\Iscr_t^{\bar{\epsilon}}
}
\geq 
\frac{K_1}{2\bar{\epsilon}^2} \frac{\log\log T}{\log T} + \frac{\epsilon'}{2 \bar{\epsilon}^2}
\right)
\\
\leq
\exp
\left(
-
\frac{T \epsilon'^2}
{8 \log^2\left((2 T + |\X|)/ p_{\min}\right) }
\right),
\end{multline*}
where $K_1$ is a constant that depends only on $|\X|$ and $|A|$.
\end{le:tvrate}
\begin{IEEEproof}
Define
\[
\Xi_t 
\defeq
\TV\left(
P(\cdot|X^{t}_{t-K+1},A^{t}_{t-K+1}),
\hat{P}_t(\cdot|X^{t}_{\tau_{c(t)}},A^{t}_{\tau_{c(t)}})
\right).
\]
We have
\begin{equation}
\label{eq:tvind}
\begin{split}
\frac{1}{T}
\sum_{t=1}^T
\Delta_t
&\geq
\frac{2\bar{\epsilon}^2}{T}
\sum_{t=1}^T
\I{
\Delta_t \geq 2 \bar{\epsilon}^2
}
\\
&\geq
\frac{2\bar{\epsilon}^2}{T}
\sum_{t=1}^T
\I{
\Xi_t
>
\bar{\epsilon}
}.
\end{split}
\end{equation}
Here, the first inequality follows by the non-negativity of Kullback-Leibler
distance. The second inequality follows from Pinsker's inequality, which
states that $\TV(\cdot,\cdot) \leq \sqrt{D(\cdot\|\cdot)/2}$.

Now, let $F_t$ be the event that the current context at time $t$,
$(X_{\tau_{c(t)}}^t,A_{\tau_{c(t)}}^{t-1})$ has never been visited in the
past. Observe that, by Lemma~\ref{le:ziv},
\begin{equation}\label{eq:Ffreq}
\sum_{t=1}^T \Inb{F_t} = c(T) \leq \frac{C_2T}{\log T}.
\end{equation}

Putting together \eqref{eq:tvind} and \eqref{eq:Ffreq} with the definition of
the event $\Iscr_{t}^{\bar{\epsilon}}$,
\[
\begin{split}
\frac{1}{T}
\sum_{t=K}^{T}
\Inb{
\Iscr_{t}^{\bar{\epsilon}}
}
& 
\leq
\frac{1}{T}
\sum_{t=1}^{T}
\left(
\I{
\Xi_t
>
\bar{\epsilon}
}
+
\Inb
{
F_t
}
\right)
\\
& 
\leq
\frac{1}{2 \bar{\epsilon}^2 T}
\sum_{t=1}^T
\Delta_t
+
\frac{C_2}{\log T}.
\end{split}
\]
Then,
\begin{multline*}
\Pr
\left(
\frac{1}{T}
\sum_{t=K}^{T}
\Inb{
\Iscr_{t}^{\bar{\epsilon}}
}
\geq
\frac{\bar{K}_1}{2\bar{\epsilon}^2} \frac{\log\log T}{\log T} + \frac{\epsilon'}{2 \bar{\epsilon}^2} + \frac{C_2}{\log T}
\right)
\\
\leq
\Pr
\left(
\frac{1}{T}
\sum_{t=1}^T
\Delta_t
\geq 
\bar{K}_1 \frac{ \log\log T}{\log T}
+
\epsilon'
\right).
\end{multline*}
By Lemma~\ref{le:contextb} and
Lemma~\ref{le:contextub}, we have
\begin{multline*}
\Pr
\left(
\frac{1}{T}
\sum_{t=1}^T
\Delta_t
\geq 
\bar{K}_1 \frac{ \log\log T}{\log T}
+
\epsilon'
\right)
\\
\leq
\exp
\left(
-
\frac{T \epsilon'^2}
{8 \log^2\left((2 T + |\X|)/ p_{\min}\right) }
\right).
\end{multline*}
This yields the desired result by defining the constant $K_1 \defeq \bar{K}_1
+ C_2/\log\log K$.
\end{IEEEproof}

\end{document}